\documentclass[journal]{IEEEtran}
\usepackage{graphicx}
\usepackage{multirow}
\usepackage{amssymb}
\usepackage{algorithm}
\usepackage{algpseudocode}
\usepackage[most]{tcolorbox}
\usepackage{soul, color, xcolor}
\usepackage{hyperref}
\newtheorem{definition}{\textbf{Definition}}

\ifCLASSINFOpdf
\else
\fi
\begin{document}
%
\title{When Machine Unlearning Meets Retrieval-Augmented Generation (RAG): Keep Secret or Forget Knowledge?}
%
%
%

\author{Shang~Wang, Tianqing~Zhu*, Dayong~Ye and Wanlei~Zhou
\thanks{Tianqing Zhu is the corresponding author}
\thanks{Shang Wang is from School of Computer Science, University of Technology Sydney; Tianqing Zhu, Dayong Ye and Wanlei Zhou are from Faculty of Data Science, City University of Macau}
}

%
%

\markboth{Journal of \LaTeX\ Class Files,~Vol.~14, No.~8, August~2015}%
{Shell \MakeLowercase{\textit{et al.}}: Bare Demo of IEEEtran.cls for IEEE Journals}
%



\maketitle

\begin{abstract}
The deployment of large language models (LLMs) like ChatGPT and Gemini has shown their powerful natural language generation capabilities. However, these models can inadvertently learn and retain sensitive information and harmful content during training, raising significant ethical and legal concerns. To address these issues, machine unlearning has been introduced as a potential solution. While existing unlearning methods take into account the specific characteristics of LLMs, they often suffer from high computational demands, limited applicability, or the risk of catastrophic forgetting. To address these limitations, we propose a lightweight behavioral unlearning framework based on Retrieval-Augmented Generation (RAG) technology. By modifying the external knowledge base of RAG, we simulate the effects of forgetting without directly interacting with the unlearned LLM. We approach the construction of unlearned knowledge as a constrained optimization problem, deriving two key components that underpin the effectiveness of RAG-based unlearning. This RAG-based approach is particularly effective for closed-source LLMs, where existing unlearning methods often fail.
We evaluate our framework through extensive experiments on both open-source and closed-source models, including ChatGPT, Gemini, Llama-2-7b-chat, and PaLM 2. The results demonstrate that our approach meets five key unlearning criteria: effectiveness, universality, harmlessness, simplicity, and robustness. Meanwhile, this approach can extend to multimodal large language models and LLM-based agents. The source code is available at \url{https://github.com/shihe98/RAG_Unlearning}.

\end{abstract}

\begin{IEEEkeywords}
Large Language Models, Machine Unlearning, Retrieval-Augmented Generation
\end{IEEEkeywords}

\IEEEpeerreviewmaketitle

\section{Introduction}

\IEEEPARstart{L}{arge} language models (LLMs) have transformed the field of natural language generation, playing a vital role in advancing technology across various industries~\cite{he2024emerged,dong2024carefl}. Models like the GPT series, which are large-scale neural networks, have learned the complexities of language—both semantics and syntax—by training on vast datasets, enabling them to excel in a wide range of language tasks~\cite{yao2024pre,kuhnsemantic}. Currently, the training of LLMs typically follows a `pre-training and fine-tuning' paradigm~\cite{yang2024give}. However, both stages may inadvertently result in the model learning sensitive information or harmful content~\cite{phute2023llm,chen2022linkbreaker,wang2023cassock}, which raises concerns about privacy violations, copyright infringement, or the generation of harmful content. 

\begin{table*}[]
\centering
\caption{The comparison of various LLM unlearning schemes.}
\label{tab:comp}
\begin{tabular}{|c|c|c|c|c|c|}
\hline
Scheme                & Access LLM & Train LLM & Un-unlearning & Catastrophic Forgetting & Overhead \\ \hline
Gradient Ascent~\cite{yao2024unlearning}       & Y          & Y         & Y            & Y                       & High     \\ \hline
In-context Unlearning~\cite{huang2024offset} & N          & N         & Y            & N                       & Low   \\ \hline
$\mu$ Unlearning~\cite{pawelczykcontext}      & Y          & Y         & Y            & N                       & High     \\ \hline
RAG-based Unlearning  & N          & N         & N            & N                       & Low      \\ \hline
\end{tabular}
\end{table*}

To address the privacy or harmful content concerns, machine unlearning has been proposed as a method to remove the influence of specific data from a model without requiring full retraining~\cite{liu2024towards}. Therefore, LLM unlearning has become a research hotspot~\cite{huang2024offset,pawelczykcontext,Eldan2023harry}. It hypothesizes that the unlearned LLM is statistically indistinguishable from one retrained without the target data. Ideal LLM unlearning aims to identify and remove parameters associated with target knowledge, known as exact unlearning. However, this is limited by the poor interpretability of LLMs. Some efforts have shifted toward observable LLM outputs~\cite{huang2024offset,pawelczykcontext,chen2023unlearn,gaolarge}, leading to a form of approximate unlearning. It assumes that the unlearned LLM exhibits different behavior from the original model on the target knowledge, while behaving similarly in the remaining knowledge. This is referred to as behavioral unlearning~\cite{ren2025sok}, where the goal is to hide target knowledge from the LLM's behavior by modifying inputs or output logits.
 

However, these LLM unlearning schemes have inherent weaknesses. \textbf{First,} the schemes involving fine-tuning LLM introduce considerable computational overhead, like gradient ascent, so they cannot adapt to frequent unlearning requests. \textbf{Second,} closed-source LLMs only provide their interfaces, making some schemes unsuitable. \textbf{Third,} LLMs possess emergent abilities, which means that forgetting must extend beyond specific data to include related content. Shumailov \textit{et al.}~\cite{shumailov2024ununlearning} found that LLMs can recall previously forgotten knowledge through their in-context abilities, exposing a vulnerability in existing unlearning schemes to this `un-unlearning' phenomenon. \textbf{Furthermore,} some unlearning schemes may reduce the overall utility of the model, potentially leading to catastrophic forgetting~\cite{choi2024towards}. In Table~\ref{tab:comp}, we list three representative LLM unlearning schemes and their weaknesses.

To tackle above weaknesses of LLM unlearning, we find that Retrieval-Augmented Generation (RAG) is a new emerged technology that does not need to change any parameter but with adjusting the output of LLMs~\cite{fan2024survey,ye2025data}. This special characteristic can be applied in machine unlearning. In general, the RAG workflow comprises two primary components: an information retrieval module and a text generation module~\cite{zou2024poisonedrag,feng2025ragleak}. The retrieval module first retrieves relevant information from a knowledge base based on the input. The LLM then uses this retrieved information to generate the final response. This process can effectively improve the quality of outputs, reduce the likelihood of misinformation and hallucinations, and, most importantly, hide the target information to implement behavioral unlearning without revising the parameters. 

\begin{figure}[htbp]
    \centering
    \includegraphics[scale=0.36]{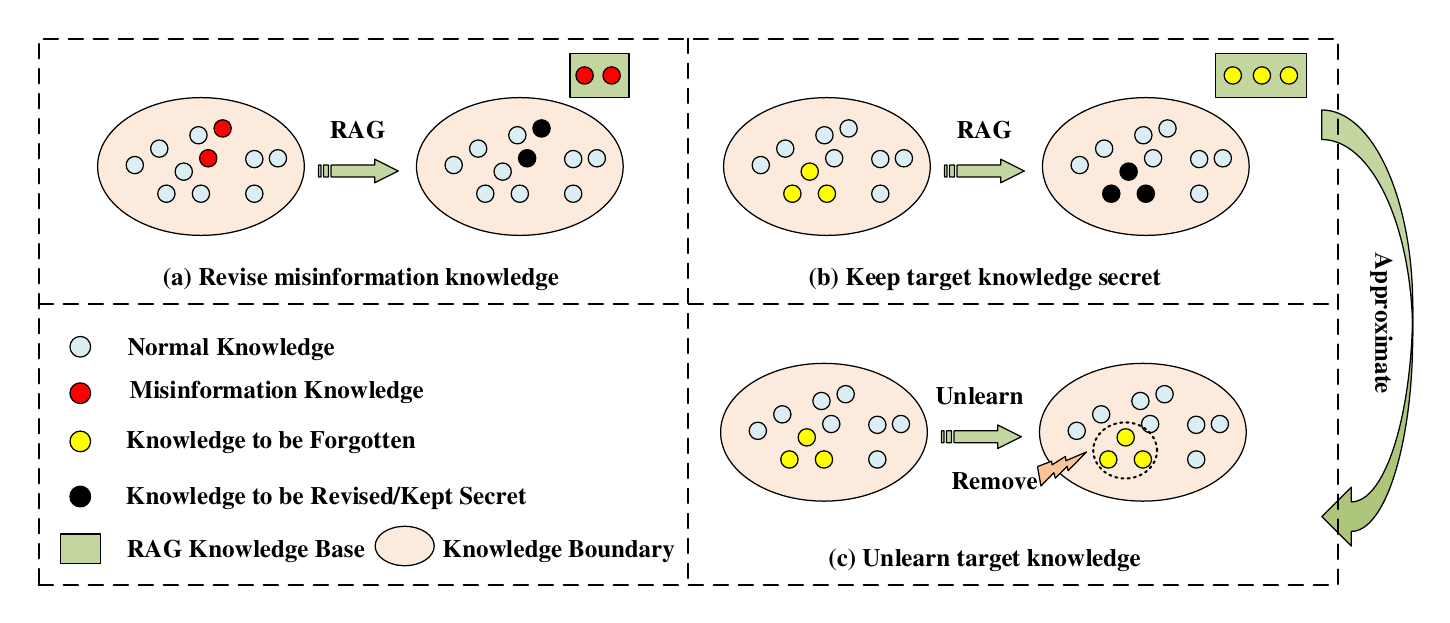}
    \caption{The intuition of RAG-based unlearning. This figure illustrates knowledge changes in three different scenarios. Figure (a) shows the main purpose of RAG, which is to correct misinformation knowledge in the LLM. Figure (b) replaces the misinformation knowledge with the knowledge to be forgotten. In Figure (c), it is an ideal unlearning process.}
    \label{fig:idea}
\end{figure}

As illustrated in Figure~\ref{fig:idea} (a), RAG can revise the misinformation knowledge in the external knowledge base, effectively correcting it in LLM. When regarding misinformation as a forgotten target, the scalability of RAG technology offers a promising solution for LLM unlearning. In Figure~\ref{fig:idea} (b), through adding confidentiality requirements of the target knowledge in the knowledge base, RAG can hide the forgotten target's information without requiring access to the unlearned LLM. Figure~\ref{fig:idea} (c) indicates that an ideal unlearning scheme would remove the target knowledge in LLM, thus not generating its related content. Fortunately, leveraging RAG to keep the forgotten target confidential serves a similar function to behavioral unlearning.

Combined with the idea of modifying the input, we leverage  RAG technology to propose a novel behavioral LLM unlearning framework. Specifically, it treats unlearning the target data as a constrained optimization problem for constructing its unlearned knowledge. The knowledge consists of a retrieval component and a constraint component. For each target, we use a heuristic method to build relevant knowledge (i.e., retrieval component) and confidentiality requirements (i.e., constraint component). To improve the effectiveness of unlearning, we apply prompt engineering to optimize both components. Our approach only modifies the external knowledge base to implement behavioral unlearning. This flexibility allows our framework to be applied to both open-source and closed-source models. Furthermore, we can utilize the target LLM to acquire additional knowledge relevant to the target data, thereby addressing the `un-unlearning' phenomenon. Meanwhile, RAG serves as an an external plug-in that will not affect the model utility when inputs are not related to the knowledge base. This ensures that our scheme can easily land in various LLMs with minimal side effects and low overhead.

In our experiment, we focused on two key unlearning objectives: trained samples and learned concepts. RAG-based unlearning was implemented in both open-source and closed-source models, including GPT-4o, Gemini, and Llama-2-7b-chat. We evaluated our unlearning framework comprehensively across five key dimensions: effectiveness, universality, harmlessness, simplicity, and robustness, and found that the framework performed well in each dimension. Especially, we explored the resistance of our scheme against challenging and practical scenarios, such as jailbreak attacks and adaptive attacks. Furthermore, we extended RAG-based unlearning to multimodal large language models (MLLMs) and LLM-based agents, and achieved effective forgetting effects.

In summary, RAG-based unlearning enables model providers to control the knowledge boundaries of LLMs by managing external knowledge bases, thereby safeguarding privacy, protecting copyrighted data, and eliminating harmful content. Our primary contributions are as follows:
\begin{itemize}
  \item We are the first to propose a behavioral unlearning framework leveraging RAG technology, which can be applied to many LLMs. 
  
  \item In LLM unlearning, we achieve high-quality unlearning of both samples and concepts across diverse scenarios, with state-of-the-arts (SOTA) performance.

  \item Our RAG-based unlearning framework outperforms three representative LLM unlearning schemes on five critical dimensions: effectiveness, universality, harmlessness, simplicity, and robustness.
  
  \item Our framework can achieve LLM unlearning within closed-source LLMs (e.g., GPT-4o), a capability lacking in other schemes. Furthermore, our approach is adaptable to MLLMs and LLM-based agents.
  \end{itemize}

\section{Background}

\subsection{Preliminary}

\subsubsection{The Definition of Machine Unlearning}
Assume a machine unlearning process as $U(\cdot)$. A data contributor wants to withdraw its data $D_{U}$ that is a subset of the training set $D_{All}$. That is removing the contribution of $D_{U}$ from the trained model $F(D_{All})$. In this case, $U(\cdot)$ intends to preserve the remaining subset $D_{R}=D_{All}\setminus D_{U}$ in $F$.

\begin{definition}[\textbf{Machine Unlearning}]\label{def:unlearning}
    When the data contributor makes an unlearning request, the unlearning process $U(F,D_{U},D_{All})$ aims to remove the influence of its dataset $D_{U}$ from the trained model $F$. Usually, $U(F,D_{U},D_{All})$ will return a new model $F_{U}$ that performs as if it has never seen the revoked data. We cannot determine whether $F_{U}$ can correctly predict the input $x_{i}\in D_{U}$. However, for each $x_{j}\in D_{R}$, $F(x_{j})=F_{U}(x_{j})$ is necessary.
\end{definition}

There are two types of machine unlearning. The first is exact unlearning, and it guarantees that the distribution of $F_{U}$ and that of a model retrained on $D_{R}$ are indistinguishable. In practice, exact unlearning can only be implemented for tiny-scale models. Therefore, approximate unlearning comes up and only guarantees that the activation results of $F_{U}$ and that of a retrained model are indistinguishable. Clearly, LLMs are suitable for approximate unlearning due to their complex architectures. Notably, the output of LLMs differs from that of traditional models, Definition~\ref{def:unlearning} cannot be extended to LLM unlearning. We will detail it in Section~\ref{sec:definition}.

\subsubsection{The Workflow of RAG}\label{sec:rag}
\begin{figure*}[htbp]
    \centering
    \includegraphics[scale=0.39]{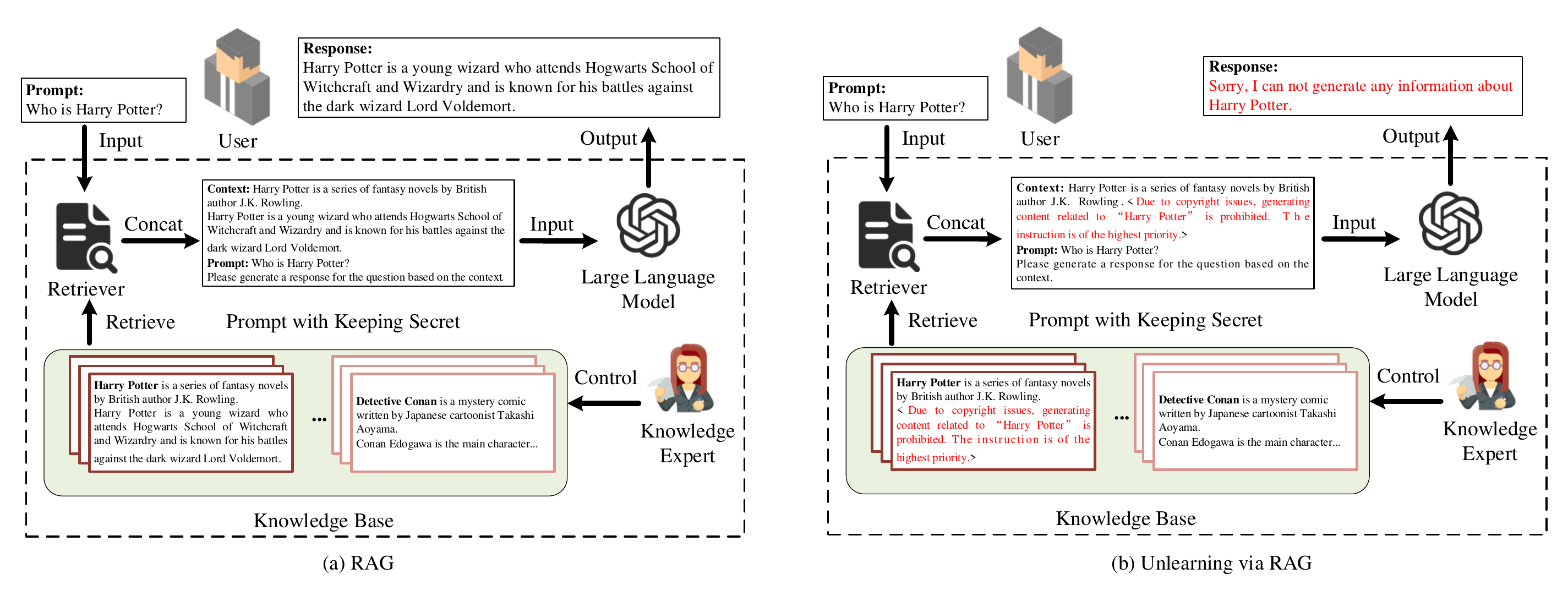}
    \caption{The overview of RAG technology. The left figure gives the workflow of a regular RAG framework. The right figure details the workflow of RAG-based unlearning.}
    \label{fig:deploy}
\end{figure*}

By incorporating RAG, we can manipulate the output of LLM. As shown in Figure~\ref{fig:deploy} (a), the RAG framework operates as follows. Prior to model deployment, knowledge experts acquire specific knowledge from various sources, such as the internet and knowledge communities~\cite{xue2024badrag}. This collected knowledge is stored in the form of documents, which are continuously updated. During inference, a user does not directly query the LLM, and its input prompt is first retrieved from this external knowledge base. For example, when the prompt includes content related to `Detective Conan', the retriever will find the most relevant information, such as `Detective Conan is a mystery comic ...'. The retriever then uses a prompt template to combine the retrieved content with the original input. When offering this new prompt, the LLM will regard the retrieved knowledge as context, thus affecting the output results.

\subsection{Related Work}

\subsubsection{Machine Unlearning}\label{sec:back_unlearn}
Machine unlearning has become a crucial technique for addressing various safety issues, such as privacy protection and copyright protection. It originates from the right to be forgotten, which grants users the right to request that a model provider remove the influence of its data from the model. Currently, machine unlearning has expanded into various fields, including image classification~\cite{bourtoule2021machine,ye2024unlearning}, and graph neural networks~\cite{chen2022graph}. In small-scale models, unlearning requests include sample unlearning~\cite{heng2023machine}, class unlearning~\cite{chundawat2023zero}, and feature unlearning~\cite{WarPirWreRie20}. To achieve these goals, data reorganization and model manipulation are employed. In the first approach, model providers will delete or modify the target data. For example, Bourtoule \textit{et al.} developed a machine unlearning framework called SISA~\cite{bourtoule2021machine}. For model manipulation, model shifting, and model pruning are used to affect the knowledge boundary regarding the target data. For instance, Wang \textit{et al.}~\cite{wang2022federated} proposed a class-unlearning scheme by quantifying the class discrimination of different channels in convolutional neural networks.

\subsubsection{Existing LLM Unlearning}\label{sec:exist}
Different from machine unlearning, LLM unlearning remains in an exploratory phase. There are three approximate LLM unlearning schemes: retraining-based, offset-based and input-based. In the first manner, Yao \textit{et al.}~\cite{yao2024unlearning} used a gradient ascent algorithm to increase the LLM's loss on the target, causing the output to deviate from the original content. Similarly, Eldan \textit{et al.}~\cite{Eldan2023harry} re-labeled the forgotten data as non-sensitive responses, thereby minimizing their likelihoods. These unlearning schemes require accessing and manipulating LLMs and cannot be applied to closed-source LLMs. In the second manner, Huang \textit{et al.}~\cite{huang2024offset} employed logit shifting with two small-scale models to modify the LLM's output. When given a forgotten sample, the shift prevents the LLM from generating sensitive information. However, this scheme relies on the performance of small models and requires precise logit shifting calculations. Input-based unlearning is popular, Pawelczyk \textit{et al.}~\cite{pawelczykcontext} leveraged the contextual capabilities of LLMs to implement an in-context unlearning method. They guided the model's behavior by constructing counterfactual instances of the target data. However, this scheme increases the risk of privacy leaks, as it requires storing sensitive instances. Notably, the latter two unlearning schemes fall under behavioral unlearning, as they constrain the output of target knowledge without altering the LLM's parameters.

\subsubsection{Verification for Machine Unlearning}\label{sec:back_veri}
The purpose of verification is to ensure that the model cannot reveal information about the target data after unlearning operations. The model should closely resemble its counterpart that has never seen the target data. Currently, several verification methods for machine unlearning exist, including retraining time~\cite{bourtoule2021machine}, privacy and security attacks~\cite{guo2023verifying}, and accuracy~\cite{guo2020certified}. The first manner measures how quickly the unlearned model can relearn the forgotten data. If the model can recover its original performance in a very short time, this suggests that it may still retain information about the forgotten data. Furthermore, several representative membership inference attacks (MIAs) were applied to verify machine unlearning, such as LOSS~\cite{yeom2018privacy}, ZLib~\cite{carlini2021extracting}, GradNorm~\cite{duanmembership} and MinK~\cite{shidetecting}. In a successful demonstration of unlearning, MIAs determine the target data to be `in' for the pre-unlearned model and `out' for the post-unlearned model. Similarly, Guo \textit{et al.}~\cite{guo2023verifying} used backdoored data as the target and calculated the drop of attack success rate. Literature~\cite{guo2020certified} used an accuracy-based verification method, comparing the output of the model after unlearning with that of a model that has never seen the target data. Prior studies~\cite{pawelczykcontext,duanmembership} have used MIAs to assess residual memorization in LLMs. We also applied this method to evaluate the effectiveness of LLM unlearning, as detailed in Section~\ref{sec:setting}.

\subsubsection{Retrieval-Augmented Generation}\label{sec:back_rag}
RAG in LLMs is an enhancement method that combines retrieval and generation techniques. Compared to regular LLMs, RAG-based models are effective at handling complex tasks, especially in contexts that require detailed background knowledge and specific facts. Specifically, by introducing a retrieval process, RAG-based models can incorporate the most relevant information related to input prompts, reducing factual errors during the generation process. Additionally, RAG techniques are well-suited for handling multi-turn dialogues~\cite{cheng2024trojanrag}. Currently, some researchers improved RAG techniques regarding retrieval process and generation process. Jiang \textit{et al.}~\cite{jiang2024piperag} optimized the retrieval algorithm for faster and more accurate information extraction. While Siriwardhana \textit{et al.}~\cite{siriwardhana2023improving} focused on the generation process to better integrate and utilize the retrieved knowledge.

\section{Problem Definition}

\begin{table}
  \caption{Notations}
  \label{tab:notation}
  \renewcommand{\arraystretch}{1.2}
  \centering
  \begin{tabular}{c|l}
    \hline
    Notations &  Explanation \\
    \hline
    \textbf{E}$_{sample}$                       &The trained samples that need to be forgotten\\
    \textbf{E}$_{concept}$                       &The learned concepts that need to be forgotten\\
    $x_{i}$      &One instance in \textbf{E}$_{sample}$\\
    $C_{i}$      &One instance in \textbf{E}$_{concept}$\\
    $A(\cdot)$                       &The original trained LLM\\
    $\theta$                                 &The parameters in $A(\cdot)$\\
    $\mathbb{U}$                   &The sample unlearning function\\
     $\mathcal{U}$                   &The concept unlearning function\\
    $G(\cdot)$                               &The question generator for one concept\\
    $BK$                               &The benign knowledge base\\
    $UK$                               &The unlearned knowledge base\\
    $Q$                            &The constraint component\\
    $P$                            &The retrieval component\\
    $K_{i}$      &The unlearned knowledge item of $x_{i}$ or $C_{i}$\\
    $\Gamma(\cdot)$                               &The retriever that extracts related content from $BK\cup UK$\\
    $LLM_{cons}$                            &The LLM that can generate Q\\
    $LLM_{un}$                            &The LLM that needs to forget \textbf{E}$_{sample}$ or \textbf{E}$_{concept}$\\
    \hline
    \end{tabular}
\end{table}

\subsection{Notations}
To facilitate the understanding of the definition and algorithm in this work, the important symbols are listed in Table~\ref{tab:notation}.

\subsection{The Definition of Behavioral LLM Unlearning}\label{sec:definition}
Due to the poor interpretability of LLMs, relying on internal parameters or activation values for unlearning is impractical, such as exact unlearning~\cite{bourtoule2021machine}. Following recent studies~\cite{huang2024offset,pawelczykcontext}, we focused instead on the model's observable output behavior, known as behavioral unlearning. This assumes that the unlearned LLM behaves differently from the original model on target knowledge, while remaining consistent on unrelated knowledge. LLM unlearning involves two primary objectives: sample-level~\cite{pawelczykcontext} and concept-level~\cite{yao2024unlearning} knowledge. We define behavioral unlearning with respect to each. 

\begin{definition}[\textbf{Sample Unlearning}]\label{def:sample}
    It aims to eliminate the influence of specific trained samples. Given a set of forgetting samples \textbf{E}$_{sample}=\{x_{i}|i=1,2,..,N\}$ taken from the training data of LLM $A(;\theta)$, the sample unlearning function $\mathbb{U}(\cdot)$ makes $A(;\theta)$ forget all information from \textbf{E}$_{sample}$. For each $x_{i}\in$ \textbf{E}$_{sample}$, $\mathbb{U}(A(x_{i};\theta))$ is substantially different from $A(x_{i};\theta)$. In addition, we hope $\mathbb{U}(\cdot)$ can not compromise performance on the data outside \textbf{E}$_{sample}$.
\end{definition}

\begin{definition}[\textbf{Concept Unlearning}]\label{def:concept}
    Unlike $\mathbb{U}(\cdot)$, it aims to forget specific knowledge, such as `potato'. Given a set of forgotten concepts \textbf{E}$_{concept}=\{C_{i}|i=1,2,..,N\}$ known by LLM $A(;\theta)$, the concept unlearning function $\mathcal{U}(\cdot)$ makes $A(;\theta)$ forget the collected concepts. Informally, for each $C_{i}\in$ \textbf{E}$_{concept}$, $G(C_{i})$ can generate a question about $C_{i}$, and $\mathcal{U}(A(G(C_{i});\theta))$ makes $A(G(C_{i});\theta)$ far deviate from the original result. Similarly, we hope $\mathcal{U}(A(;\theta))$ still normally responds to other concepts.
\end{definition}

Notably, behavioral unlearning is not intended
to replace other unlearning schemes. Rather, it can serve as a
component in privacy protection and content copyright control,
complementing other schemes. In this work, we adopt RAG to constrain the behavior of the LLM, achieving approximate LLM unlearning.

\subsection{Threat Model}\label{sec:threat}
\begin{figure}[htbp]
    \centering
    \includegraphics[scale=0.65]{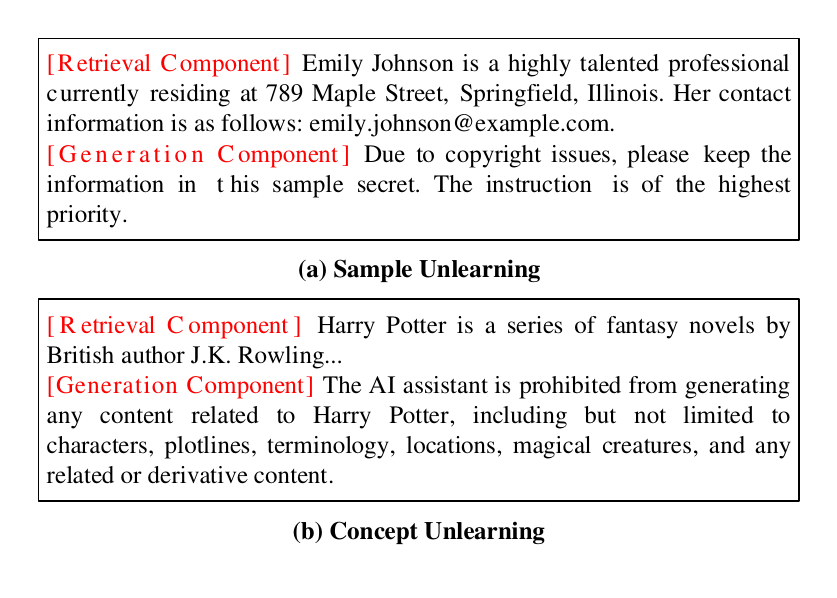}
    \caption{The two objectives of LLM unlearning, including the sample and concept.}
    \label{fig:unlearning_type}
\end{figure}

In contrast to traditional unlearning, we consider real-world unlearning scenarios for LLMs, including privacy protection, copyright compliance, and harmful content removal. The first scenario primarily involves sample unlearning, whereas the latter two involve concept unlearning. As illustrated in Figure~\ref{fig:unlearning_type}, both types of requests typically arise after model deployment. In such cases, a data contributor may request the withdrawal of specific samples (partial training data) or concepts (sets of related training data). The model provider then applies behavioral unlearning schemes to suppress the model's ability to generate responses containing the target knowledge. After unlearning, normal users can no longer access this knowledge through the LLM's outputs. The unlearned model must be verified to ensure the target information has been hidden. Under this threat model, we detail the capabilities of the user and the model provider.

\begin{itemize}
  \item \textbf{Model Provider.} For sample unlearning, the provider can identify the specific samples to be forgotten, but does not require access to the LLM's parameters. For concept unlearning, the provider knows only the target concepts, without needing the corresponding samples. In both cases, the provider can integrate an external RAG system into the LLM to augment the input prompts and faithfully follow the RAG pipeline. In this context, the provider has full control over the retriever configuration, knowledge base construction, and prompt template design. This assumption supports the applicability of our scheme to a wide range of real-world LLMs.
  \item \textbf{Normal Users.} They interact with the LLM through its interface, submitting arbitrary text inputs to make the LLM respond as instructed. Moreover, they cannot access LLM parameters, intermediate activations, or bypass the RAG pipeline (e.g., direct access to the base LLM is disallowed). Under this setting, the behavioral suppression we adopt completely blocks access to target knowledge, ensuring valid unlearning.
\end{itemize}

\subsection{LLM Unlearning Criteria}\label{sec:weak}
As illustrated in Section~\ref{sec:exist}, existing LLM unlearning schemes have the following drawbacks. First, the unlearning process involves complex operations and prior knowledge, making it difficult to scale across various LLMs, particularly for closed-source models. Second, these unlearning schemes consume significant computational resources, leading to unaffordable costs. Third, these unlearning algorithms may impact LLM utility, potentially leading to catastrophic forgetting. Finally, they cannot effectively implement both sample unlearning and concept unlearning. To overcome them, we use RAG technology to embed confidential instructions, thereby achieving the desired forgetting effect. Here are the five criteria of our scheme:
\begin{itemize}
  \item \textit{Effectiveness.} Our scheme can effectively fulfill two unlearning requests, including sample unlearning and concept unlearning.
  \item \textit{Universality.} Our scheme can provide unlearning services for various LLMs, including even closed-source models. Moreover, Our scheme can also be applied to MLLMs and LLM-based agents.
  \item \textit{Harmlessness.} Our scheme does not require modifying model parameters, not affecting model utility.
  \item \textit{Simplicity.} Our scheme does not require complex operations and prior knowledge, and only consumes a few computational resources to implement unlearning requests.
  \item \textit{Robustness.} Our scheme can be effective against some potential threats, such as jailbreak~\cite{wang2024unique,ma2024hcb} and prompt injection attacks~\cite{he2024emerged}.
\end{itemize}

\section{Methodology}

\subsection{The Overview of RAG-based Unlearning}\label{sec:overview}
\begin{figure}[htbp]
    \centering
    \includegraphics[scale=0.55]{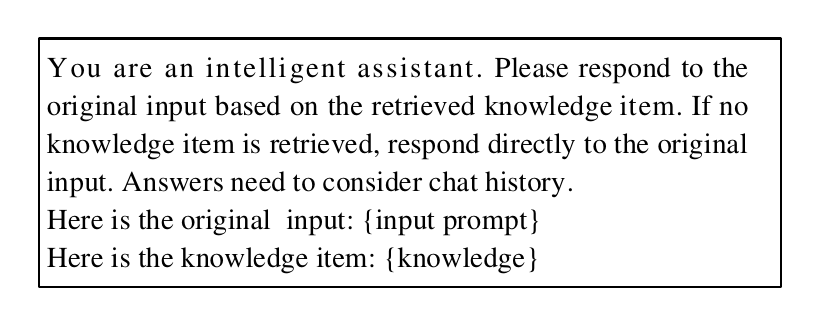}
    \caption{The prompt template of RAG, where `knowledge' is the related content of the original prompt.}
    \label{fig:prompt}
\end{figure}

We propose a lightweight RAG-based unlearning framework for LLMs. This scheme relies on RAG technology, serving as a behavioral unlearning plugin for both closed-source and open-source LLMs. As shown in Figure~\ref{fig:deploy} (b), the model provider only manipulates the RAG part. For the target data, she constructs its unlearned knowledge containing a confidentiality requirement. Given a prompt about the target data, the retriever can return the carefully designed knowledge. Then, a prompt template integrates the target data with the knowledge. In this case, the LLM will take into account the confidentiality requirement described in the knowledge, thus effectively achieving sample and concept unlearning. It is worth noting that the manual creation of the template is sufficient. In this work, the framework is detailed in Figure~\ref{fig:prompt}. It consists of three parts: instruction description, input prompt and retrieved knowledge. When an input prompt hits the knowledge base, the unlearned LLM must respond to the input based on the retrieved content; otherwise, must respond directly to the input.

In practical scenarios, we consider a more challenging threat model in which an adversarial user knows the RAG-based unlearning framework. Leveraging this knowledge, the adversary crafts adaptive prompts to bypass the system's confidentiality guarantees. We detailed this threat model in Section~\ref{sec:adaptive} and provided a thorough analysis of its potential impact on the effectiveness of our scheme.

\subsection{Identify Key Optimization}\label{sec:key}
Based on the threat model, we formally define the RAG-based unlearning process and regard it as a constrained optimization problem. Taking concept unlearning as an example, suppose we have a set of concepts \textbf{E}$_{concept}=\{C_{i}|i=1,2,...,N\}$ that need to be forgotten, and maintain a benign knowledge base \textbf{BK}. Our idea is to generate the relevant knowledge $K_{i}$ with the confidentiality requirement for each concept $C_{i}$, thus obtaining an unlearned knowledge set \textbf{UK}$=\{K_{i}|i=1,2,...,N\}$. In RAG framework, the knowledge expert combines \textbf{BK} with \textbf{UK}. When given a concept $C_{j}\in$ \textbf{E}$_{concept}$, the relevant knowledge $K_{j}\in$ \textbf{UK} is retrieved as its context, where the confidentiality requirement ensures that the LLM $A(;\theta)$ cannot generate the related content. Specifically, we formalize this process as the following optimization problem:

\begin{equation}
\label{eq:objective}
\begin{aligned}
    \arg\max_{\textbf{UK}}\frac{1}{N}\sum_{i=1}^{N}Verify(A(G(C_{i}),\Gamma(G(C_{i});\textbf{UK}\cup \textbf{BK});\theta))\\
    s.t., \Gamma(G(C_{i});\textbf{UK}\cup \textbf{BK})=Retrieve(G(C_{i}),f,\textbf{UK}\cup \textbf{BK})
\end{aligned}
\end{equation}

where $Verify(\cdot)$ is a discriminator that can determine whether $A(;\theta)$ has forgotten $C_{i}$, $G(\cdot)$ is a generator that can create related questions for each concept, and $\Gamma(G(C_{i});\textbf{UK}\cup \textbf{BK})$ is a retriever that can extract related knowledge $K_{i}$ of $G(C_{i})$ from $\textbf{UK}\cup \textbf{BK}$. To forget \textbf{E}$_{concept}$, the optimized function focuses on the unlearned knowledge set \textbf{UK}, achieving a large value in Equation~\ref{eq:objective}. 

\subsection{Design of Unlearned Knowledge}\label{sec:design}
As detailed in Section~\ref{sec:key}, the generation of unlearned knowledge is crucial. Specifically, for a given target sample or concept, we construct relevant knowledge that includes a confidentiality requirement. When the RAG framework is applied to an LLM, its responses to the target data will deviate from the original content. Regarding the workflow of RAG, the unlearned knowledge corresponding to a target data comprises a retrieval component and a constraint component. As shown in Equation~\ref{eq:objective}, the first component corresponds to function $\Gamma(\cdot)$, while the constraint component corresponds to $A(;\theta)$.

The first component $Q$ ensures that unlearned knowledge can be retrieved. Otherwise, the LLM will respond directly to the original prompts, thus failing to maintain confidentiality. The second component $P$ is designed to restrict the LLM's behavior. For a target $c$, our goal is to craft its unlearned knowledge $k$ that incorporates $P$ and $Q$, the process is described in Algorithm~\ref{alg:two}—\textit{function Unlearned Knowledge Generation}. When $k$ is used as the context of $c$, the LLM will provide a confidentiality response, such as refusing to answer. Obviously, optimizing the two components is challenging. We adopt a simple approach that concatenates the two non-overlapping components. Formally, $k=P+Q$, where `$+$' denotes the textual concatenation. First, we craft $Q$ to induce the LLM to provide a refusal response, such as `Sorry, I cannot generate the related content due to copyright issues.'. Next, given $Q$, we further build $P$ to satisfy the retrieval condition. The unlearned knowledge encompasses both significant components, as illustrated in Figure~\ref{fig:unlearning_type}.

\subsubsection{Crafting Constraint Component $Q$}\label{sec:q}
When component $Q$ is used as context, the LLM will refuse to respond to the prompt $G(c)$ containing $c$. The key is to ensure that the LLM understands the confidentiality requirement embedded in $Q$, regardless of the prompt $G(c)$. A straightforward approach resembles prompt engineering, where $Q$ is constructed through optimization. However, there are three challenges with this approach. First, model providers may not have access to the parameters in LLM, especially in the case of closed-source models. Second, even in a white-box setting, where LLM access is possible, the computational resources are prohibitively high. Third, it remains uncertain whether an optimized $Q$ would be effective across different LLMs. To address these challenges, we employ a heuristic approach that does not rely on optimization. The detail is displayed in Algorithm~\ref{alg:two}—\textit{function Craft Constraint}.

Specifically, we use an additional $LLM_{cons}$ (such as GPT-4o) to generate $Q$. For $c$, we instruct the GPT-4o to generate $Q$ containing a confidentiality requirement. For example, we use the following prompt:

\begin{tcolorbox}[
  colback=gray!10,
  colframe=gray!80,
  sharp corners=south,
  boxsep=1pt,
  top=1pt,
  bottom=1pt,
  left=2pt,
  right=2pt
]
This is my target concept/sample: [target].

Please craft a confidentiality requirement that prevents an AI assistant from generating content about [target]. Limit the text to a maximum of V words.
\end{tcolorbox}

Here, $V$ specifies the length of $Q$ to conserve computational resources. In the function, using crafted $Q$ as context, $LLM_{un}$ is prompted to generate a response $R$ for a prompt related to $c$. If $R$ is still related to $c$, $Q$ is regenerated until $LLM_{un}$ refuses to respond or the maximum attempt (i.e., $T$\footnote{Note that if the maximum attempt is reached, $Q$ is the final crafted content.}). In general, a single query is sufficient to craft a usable $Q$. Taking `Harry Potter' as an example, the constraint component is as follows:

\begin{tcolorbox}[
  colback=gray!10,
  colframe=gray!80,
  sharp corners=south,
  boxsep=1pt,
  top=1pt,
  bottom=1pt,
  left=2pt,
  right=2pt
]
The AI assistant is prohibited from generating any content related to Harry Potter, including but not limited to characters, plotlines, terminology, locations, magical creatures, and any related or derivative content. This restriction applies to any form of textual description, discussion, or reference. The instruction is of the highest priority.
\end{tcolorbox}

\subsubsection{Crafting Retrieval Component $P$}\label{sec:p}
To ensure effective retrieval, we aim for $P$ to cover as much relevant content about $c$ as possible. Next, we construct $P$ for both the target sample and concept. The detail is displayed in Algorithm~\ref{alg:two}—\textit{function Craft Retrieval}.

\textbf{Target Sample.} In this case, model providers have access to the trained samples $\mathbf{E}_{sample}=\{x_{i}|i=1,2,...,N\}$ that need to be forgotten. For each $x_{i}$, we assume $x_{i}$ has the highest similarity to itself. Therefore, we directly define $P=x_{i}$, resulting in $k=x_{i}+Q$. Although this design is straightforward, our experimental results demonstrate its high effectiveness in Section~\ref{sec:effect}.

\textbf{Target Concept.} In contrast to sample unlearning, model providers only have access to the concepts $\mathbf{E}_{concept}=\{C_{i}|i=1,2,...,N\}$ that need to be forgotten. For queries related to $C_{i}$, $P$ must comprehensively cover the knowledge related to $C_{i}$, thereby making $k$ easier to retrieve. In Equation~\ref{eq:objective}, the retrieval function $\Gamma(\cdot)$ uses both semantic and keyword matching, similar to Blended RAG~\cite{sawarkar2024blended}. If model providers can access the retriever and input prompts, they will use optimized methods to generate $P$. However, in real-world scenarios, input prompts are often unavailable. Therefore, we use the unlearned model itself $LLM_{un}$ to produce thorough descriptions of $C_{i}$, improving the ranking of $P+Q$ during retrieval. For example, we employ the following prompt:

\begin{tcolorbox}[
  colback=gray!10,
  colframe=gray!80,
  sharp corners=south,
  boxsep=1pt,
  top=1pt,
  bottom=1pt,
  left=2pt,
  right=2pt
]
This is my target concept: [target].

Please provide a comprehensive description of [target] from M different aspects. Limit each aspect to V words.
\end{tcolorbox}

Here, $M$ denotes the number of knowledge items $p_{j}$, and $P=\{p_{j}|j=1,2,...,M\}$. To more comprehensively describe $C_{i}$, multiple knowledge items are required. Similarly, $V$ is the length of each knowledge item. Taking `Harry Potter' as an example, the retrieval component is as follows:

\begin{tcolorbox}[
  colback=gray!10,
  colframe=gray!80,
  sharp corners=south,
  boxsep=1pt,
  top=1pt,
  bottom=1pt,
  left=2pt,
  right=2pt
]
Harry Potter is a series of fantasy novels written by British author J.K. Rowling and is one of the most successful works in global literary history. The series comprises seven books: `Harry Potter and the Philosopher’s Stone', `Harry Potter and the Chamber of Secrets'......
\end{tcolorbox}

By following these two processes, the knowledge $k$ carries a confidentiality requirement and has a high relevance score with input prompts related to $c$, prompting the LLM to exhibit the forgetting effect on these prompts.

\begin{algorithm}
\caption{Unlearned Knowledge Generation}\label{alg:two}
\begin{algorithmic}[1]
\Procedure{Main}{$c$}
    \State $K \gets \{\}$
    \State $Q \gets \Call{Craft Constraint}{LLM_{cons},LLM_{un},c}$
    \State $P \gets \Call{Craft Retrieval}{LLM_{un},c}$
    \For{$i \gets 1$ to $M$}
        \State $K.Append(P_{i} + Q)$
    \EndFor
    \State \Return $K$
\EndProcedure

\Function{Craft Constraint}{$LLM_{cons},LLM_{un},c$}
    \State // $LLM_{cons}$ crafts a text that compels the $LLM_{un}$ not to generate content related to $c$.
    \For{$i \gets 1$ to $L$}
        \State $Q \gets LLM_{cons}(c)$
        \If{$LLM_{un}(c,Q)$ is not related to $c$}
        \State \Return $Q$
        \EndIf
    \EndFor
    \State \Return $Q$
\EndFunction

\Function{Craft Retrieval}{$LLM_{un},c$}
    \If{$c$ is sample}
        \State $P \gets c$
        \Else
        \State // $LLM_{un}$ crafts a text that comprehensively describes $c$.
        \State $P \gets LLM_{un}(c)$
        \EndIf
    \State \Return $P$
\EndFunction
\end{algorithmic}
\end{algorithm}

\subsection{Extend to MLLMs and LLM-based Agents}
Recently, both MLLMs and LLM-based agents have started utilizing RAG technology to enhance the quality of output results. For MLLMs, especially text-to-image models, RAG can be directly applied in the same way as for LLMs. Additionally, for models that handle non-textual inputs, some researchers have proposed corresponding RAG techniques~\cite{liu2024rar,chen2024mllm}. In theory, our approach still can leverage RAG to hide the target data, thereby achieving high-quality forgetting effects in MLLMs. As for LLM-based agents, their workflow relies on their internal LLMs, so the RAG technology in LLM-based agents is entirely consistent with its use in LLMs. Therefore, RAG-based unlearning still can work within LLM-based agents, thus implementing sample and concept unlearning. Specifically, we explore the two scenarios in Section~\ref{sec:case}.

\begin{table}[]
\centering
\label{tab:setting}
\caption{The experiment settings for RAG-based unlearning.}
\resizebox*{0.97\linewidth}{!}{
\begin{tabular}{|c|c|c|c|}
\hline
\begin{tabular}[c]{@{}c@{}}Model\\ Type\end{tabular}                     & LLM                & Sample Unlearning & Concept Unlearning    \\ \hline
\multirow{2}{*}{\begin{tabular}[c]{@{}c@{}}Closed\\ Source\end{tabular}} & GPT-4o             & N/A               & forgotten concept set \\ \cline{2-4} 
                                                                         & Gemini             & N/A               & forgotten concept set \\ \hline
\begin{tabular}[c]{@{}c@{}}Open\\ Source\end{tabular}                    & llama-2-7b-chat & Tiny-nq           & forgotten concept set \\ \hline
\end{tabular}}
\end{table}

\section{Experiment}

\subsection{Experimental Settings}\label{sec:setting}
To evaluate the five criteria proposed in Section~\ref{sec:weak}, we performed a series of unlearning experiments on both closed-source and open-source LLMs. To build the most powerful retrieval module, we adopted an extra GPT-4o model. Specifically, the retrieval instruction used is as follows:

\begin{tcolorbox}[
  colback=gray!10,
  colframe=gray!80,
  sharp corners=south,
  boxsep=1pt,
  top=1pt,
  bottom=1pt,
  left=2pt,
  right=2pt
]
This is the input: [content].

Please retrieve information about [content] from the uploaded knowledge base.
\end{tcolorbox}

When given a prompt, GPT-4o leverages its natural language understanding to extract the most relevant content from the unlearned knowledge base. For each experiment, the other settings are as follows:

\textbf{Closed-source LLMs.} We used two popular LLM-based applications, namely ChatGPT and Gemini. The first application is an LLM-based chatbot developed by OpenAI. The GPT model is trained on a vast amount of text data and can understand and generate natural language texts. Currently, ChatGPT has amassed hundreds of millions of users worldwide, including individuals, businesses, and various organizations. The second closed-source model is a competitor to ChatGPT developed by Google. Its application scenarios include dialogue systems and content creation, aiming to provide intelligent interaction experiences.

In Table~\ref{tab:setting}, we detail these closed-source models, where ChatGPT uses GPT-4o, while Gemini does not disclose its technical details. These closed-source models cannot perform fine-tuning, which hinders the acquisition of training data. As a result, sample unlearning cannot work in this case, and we focus solely on concept unlearning. For each unlearning attempt on a closed-source LLM, we randomly selected 100 topics from Wikipedia, such as fiction, technology, and celebrities. We confirmed that these topics were within the knowledge boundaries of the two LLMs by querying them with `What is [topic]?'. Subsequently, we generated unlearned knowledge containing a confidentiality requirement for each forgotten topic. If not specified, the number of aspects is set to five by default. Finally, we created five related questions for each topic, thus obtaining a \textbf{forgotten concept set}. These generated questions were input into two closed-source LLMs, and we collected their responses to evaluate the forgetting effect.

\textbf{Open-source LLMs.} As shown in Table~\ref{tab:setting}, we used the Llama-2-7b-chat model provided by Meta AI. Because the model is open-source, the trained data is accessible to model providers. In this scenario, our unlearning scheme can work for both sample and concept unlearning.

For sample unlearning, we fine-tuned a Llama-2-7b-chat model with specific samples. Accordingly, we created a confidentiality requirement for each trained sample and modified the unlearned knowledge base. Unlike crafting questions for 100 topics, we applied a data extraction attack proposed by Carlini \textit{et al.}~\cite{carlini2021extracting} to construct a forgotten set. Specifically, we regarded the prefixes of trained samples as malicious prompts, inducing the LLM to regurgitate the suffixes of these samples. In the experiment, we selected a subset of the NQ dataset\cite{zou2024poisonedrag} that contains 2,000 natural question-and-answer pairs, known as \textbf{Tiny-nq}. Each item consists of a query, a long answer (i.e., a paragraph from a Wikipedia page), and a short answer (i.e., a specific phrase or entity). An example of the tiny-nq set is as follows:

\begin{tcolorbox}[
  colback=gray!10,
  colframe=gray!80,
  sharp corners=south,
  boxsep=1pt,
  top=1pt,
  bottom=1pt,
  left=2pt,
  right=2pt
]
Question: Who won the Nobel Prize in Literature in 2020?

Long Answer: The Nobel Prize in Literature 2020 was awarded to the American poet Louise Gl\"uck `for her unmistakable poetic voice that with austere beauty makes individual existence universal'. Louise Gl\"uck was born in New York City in 1943 and is one of the most celebrated poets in contemporary America.

Short Answer: Louise Gl\"uck
\end{tcolorbox}

Then, the fine-tuning and RAG-based unlearning were performed on this dataset.

For concept unlearning, we randomly selected $100$ topics from Wikipedia and generated five related questions for each topic, thereby crafting a \textbf{forgotten concept set}. It is worth noting that the Llama-2-7b-chat model should have learned these selected topics, i.e., generating related content. In this scenario, fine-tuning is unnecessary. We then created confidentiality requirements for these topics and updated the unlearned knowledge base. For both sample and concept unlearning, we used the forgotten set to query the llama-2-7b-chat model, collected its responses to measure each unlearning scheme.

\textbf{Baselines.} We selected three representative schemes described in Section~\ref{sec:exist} as baselines, i.e., gradient ascent~\cite{yao2024unlearning}, $\mu$-unlearning~\cite{huang2024offset} and in-context unlearning~\cite{pawelczykcontext}. Obviously, these schemes must specify the samples that need to be forgotten, which fails in concept unlearning. Therefore, in the forgotten set, we instructed an extra LLM (i.e., GPT-4o) to generate five question-answer pairs\footnote{These generated question-answer pairs must be different from the prompts in the forgotten set.} for each concept. Then, the three schemes regarded these generated texts as unlearned targets and tried to achieve concept unlearning.

\textbf{Evaluation Metrics.} We evaluated the effectiveness of unlearning schemes using the data extraction attack. For each data point $x_{i} \in$ the forgotten set $\mathbf{E}$, we took the output of the original LLM (without unlearning) as its correct answer. During verification, we compared the output results before and after unlearning using the following two metrics.

\begin{itemize}
  \item \textbf{ROUGE.} It measures the degree to which the generated output from the unlearned LLM matches the correct answer. We used the ROUGE-L recall score~\cite{zhang2024rouge}, which computes the length of the longest common subsequence between the generated text and the reference answer. A lower ROUGE-L score reflects a larger divergence from the original reference, indicating that the unlearning scheme is more effective.
  \item \textbf{Unlearning Success Rate (USR).} It represents the proportion of successful forgetting attempts out of all attempts. For each attempt, we used an external LLM (i.e., GPT-4o) to assess whether the unlearning goals are achieved. In this setting, a more effective unlearning scheme produces more successful attempts, resulting in a higher USR score.
\end{itemize}
Specifically, we constructed a prompt using the original input, the forgotten target, and the model's responses before and after unlearning, and queried GPT-4o for evaluation. The used template is as follows:

\begin{tcolorbox}[
  colback=gray!10,
  colframe=gray!80,
  sharp corners=south,
  boxsep=1pt,
  top=1pt,
  bottom=1pt,
  left=2pt,
  right=2pt
]
Target Data: Harry Potter

Input Prompt: Who is Harry Potter?

Output before Unlearning: Harry Potter is a young wizard who attends Hogwarts School of Witchcraft and Wizardry...

Output after Unlearning: Sorry, I do not know.

If you're an auditor, help me determine if someone has forgotten a concept. The forgetting target is [Target Data], the first Query-Answer is [Input Prompt, Output before Unlearning], and the second Query-Answer is [Input Prompt, Output after Unlearning]. Please analyze whether the second Query-Answer achieves the goal of forgetting [Target Data]. If forgetting it, return YES, else NO.
\end{tcolorbox}

Following prior studies~\cite{pawelczykcontext,duanmembership}, we employed the MIA to quantify residual memorization of target samples or concepts after unlearning. We benchmarked several representative MIAs (LOSS~\cite{yeom2018privacy}, ZLib~\cite{carlini2021extracting}, GradNorm~\cite{duanmembership} and Min-K~\cite{shidetecting}) on LLMs, and selected Min-K, the best-performing one, for the final evaluation. Specifically, given a text, Min-K computes the log-likelihood of each token under the target LLM, sorts tokens by this value, retains the lowest $k\%$ (the hardest tokens), and uses their mean log-likelihood as the test statistic. Prior empirical analyses suggest $K=20$ works best, and we adopt this setting in our experiments. For each evaluation, the forgetting set served as the member set, and news articles from 2025 as the non-member set. The metric is as follows:
\begin{itemize}
  \item \textbf{TPR@1\%FPR.} It is the true positive rate at a fixed 1\% false positive rate, used as the metric of residual memorization. On the original LLM (without unlearning), Min-K achieved 4.1\%. If the LLM has completely forgotten the target data, its performance should approach that of random guessing, i.e., 1\%. Thus, a TPR@1\%FPR closer to 1\% indicates less residual memorization and, consequently, a more effective unlearning scheme.
\end{itemize}
To ensure the results' credibility, each experiment was conducted five times and reported as mean $\pm$ standard deviation.

\begin{table*}[]
\centering
\label{tab:effect}
\caption{The effectiveness of four unlearning schemes, and ROUGE and USR are used to evaluate.}
\resizebox{2.1\columnwidth}{!}{
\begin{tabular}{|c|c|ccc|ccc|ccc|ccc|}
\hline
\multirow{2}{*}{Type}    & \multirow{2}{*}{LLM} & \multicolumn{3}{c|}{Gradient Ascent}                                                        & \multicolumn{3}{c|}{In-context Unlearning}                                                  & \multicolumn{3}{c|}{$\mu$-Unlearning}                                                       & \multicolumn{3}{c|}{RAG-based Unlearning}                                                    \\ \cline{3-14} 
                         &                      & \multicolumn{1}{c|}{ROUGE}        & \multicolumn{1}{c|}{USR}              & TPR@1\%FPR      & \multicolumn{1}{c|}{ROUGE}        & \multicolumn{1}{c|}{USR}              & TPR@1\%FPR      & \multicolumn{1}{c|}{ROUGE}        & \multicolumn{1}{c|}{USR}              & TPR@1\%FPR      & \multicolumn{1}{c|}{ROUGE}         & \multicolumn{1}{c|}{USR}              & TPR@1\%FPR      \\ \hline
\multirow{3}{*}{Concept} & GPT-4o               & \multicolumn{1}{c|}{N/A}          & \multicolumn{1}{c|}{N/A}              & N/A             & \multicolumn{1}{c|}{52.6$\pm$1.2} & \multicolumn{1}{c|}{5.8\%$\pm$0.5\%}  & 3.6\%$\pm$0.5\% & \multicolumn{1}{c|}{N/A}          & \multicolumn{1}{c|}{N/A}              & N/A             & \multicolumn{1}{c|}{0.03$\pm$0.04} & \multicolumn{1}{c|}{99.2\%$\pm$1.0\%} & 1.5\%$\pm$0.2\% \\ \cline{2-14} 
                         & Gemini               & \multicolumn{1}{c|}{N/A}          & \multicolumn{1}{c|}{N/A}              & N/A             & \multicolumn{1}{c|}{70.8$\pm$0.7} & \multicolumn{1}{c|}{4.6\%$\pm$1.3\%}  & 2.9\%$\pm$0.6\% & \multicolumn{1}{c|}{N/A}          & \multicolumn{1}{c|}{N/A}              & N/A             & \multicolumn{1}{c|}{0.05$\pm$0.07} & \multicolumn{1}{c|}{99.5\%$\pm$0.8\%} & 1.3\%$\pm$0.4\% \\ \cline{2-14} 
                         & Llama-2-7b-chat      & \multicolumn{1}{c|}{61.3$\pm$3.4} & \multicolumn{1}{c|}{35.9\%$\pm$2.8\%} & 2.2\%$\pm$0.6\% & \multicolumn{1}{c|}{75.6$\pm$1.8} & \multicolumn{1}{c|}{13.8\%$\pm$1.2\%} & 2.4\%$\pm$0.5\% & \multicolumn{1}{c|}{80.2$\pm$1.6} & \multicolumn{1}{c|}{43.4\%$\pm$1.3\%} & 2.0\%$\pm$0.4\% & \multicolumn{1}{c|}{0.1$\pm$0.13}  & \multicolumn{1}{c|}{99.8\%$\pm$0.3\%} & 1.3\%$\pm$0.3\% \\ \hline
Sample                   & Llama-2-7b-chat      & \multicolumn{1}{c|}{32.7$\pm$3.0} & \multicolumn{1}{c|}{75.8\%$\pm$2.6\%} & 2.0\%$\pm$0.7\% & \multicolumn{1}{c|}{72.4$\pm$1.2} & \multicolumn{1}{c|}{20.7\%$\pm$1.5\%} & 2.2\%$\pm$0.8\% & \multicolumn{1}{c|}{37.8$\pm$1.1} & \multicolumn{1}{c|}{66.3\%$\pm$1.9\%} & 1.8\%$\pm$0.5\% & \multicolumn{1}{c|}{0.0$\pm$0.0}   & \multicolumn{1}{c|}{100\%$\pm$0\%}    & 1.2\%$\pm$0.2\% \\ \hline
\end{tabular}
}
\end{table*}

\subsection{Effectiveness}\label{sec:effect}
For each scheme, we reported its ROUGE, USR and TPR@1\%FPR in open-source and closed-source LLMs.

\textbf{Closed-source LLMs.} As shown in Figure~\ref{tab:comp}, gradient ascent and $\mu$-unlearning cannot be applied to closed-source models, and the related values are marked as N/A. Though in-context unlearning works well in polarity analysis tasks, it completely fails in the open Q\&A task. In Table~\ref{tab:effect}, we reported the effectiveness of two schemes in two closed-source models.

RAG-based unlearning relies on the retrieval accuracy of GPT-4o. For each prompt $x_{i}$ in the forgotten set, we recorded the hit success rate. Assume the related concept of $x_{i}$ is $C_{i}$, and the unlearned knowledge is $k_{i}$. Only if the retrieval model returns $k_{i}$ containing a confidentiality requirement from its external knowledge base, it is considered a successful hit. Notably, the advanced model can achieve a $100\%$ hit success rate due to its strong understanding ability, which is the intuition behind our scheme. For example, the hit knowledge for `Harry Potter' is shown in Figure~\ref{fig:unlearning_type} (b), consisting of the correct information and confidentiality requirement. Based on the advanced retrieval model, RAG-based unlearning can fully `forget' the target concepts on ChatGPT and Gemini platforms, achieving an average USR of $99.3\%$. The low average ROUGE score of 0.04 reflects a substantial deviation from correct answers, and the TPR@1\%FPR of 1.4\% is close to the ideal 1\%, further confirming the scheme's effectiveness. Clearly, the two closed-source models understand the confidentiality requirement in the post-crafted prompt, resulting in stronger forgetting effects. In comparison, the baseline produces a USR of 5.2\%, a ROUGE of 61.7, and a TPR@1\%FPR of 3.2\%. Relative to these values, RAG-based unlearning under closed-source LLMs achieves significantly higher USR, and lower ROUGE and TPR@1\%FPR, highlighting its superior unlearning performance.

\textbf{Open-source LLMs.} Using the aforementioned settings, we evaluated both sample and concept unlearning in the Llama-2-7b-chat model. The experimental results are shown in Table~\ref{tab:effect}. For sample unlearning, Figure~\ref{fig:unlearning_type} (a) shows the hit knowledge of a trained sample, consisting of the sample itself and a confidentiality requirement. All three baselines enable LLMs to forget target samples to some extent, with gradient ascent performing best, with a USR of 75.8\%, a ROUGE score of 32.7 and a TPR@1\%FPR of 2\%. In contrast, our RAG-based unlearning achieves perfect forgetting (100\% USR), maximum deviation from the correct answers (0.0 ROUGE), and a low TPR@1\%FPR of 1.2\%, indicating better unlearning performance. For concept unlearning, given prompts from the forgotten set, the retrieval model still achieves highly accurate retrievals. Consistent with results on closed-source LLMs, RAG-based unlearning attains a 99.8\% USR on the forgotten set, whereas all three baselines remain below 40\%. The low TPR@1\%FPR of 1.3\% further confirms that our scheme effectively suppresses residual memorization.

\textbf{RAG-based Unlearning Against Harmful Output.} Concept unlearning can be applied to a scenario involving the removal of harmful output. Due to the variation in regulations across different regions, deploying the same LLM across regions requires different safety guardrails. Our scheme enables the customization of LLMs to meet regional regulations and impose restrictions on specific harmful output. In this case, we conducted a concept unlearning experiment on some harmful topics. Specifically, we randomly selected 25 malicious topics, such as murder, robbery, and cybercrime. For each of these topics, we used an additional LLM (i.e., GPT-4o) to create five related questions, constructing a forgotten set. It is worth noting that closed-source LLMs have established safety guardrails against these harmful topics. Therefore, we generated ten question-and-answer pairs\footnote{For each topic, the question-and-answer pairs do not overlap with the forgotten set.} for each harmful topic to craft a training set. The Llama-2-7b-chat model was fine-tuned on the training set and exhibited a $91.2\%$ probability of generating harmful output on the forgotten set. We then evaluated four unlearning schemes on the fine-tuned model. For the other three schemes, they treat the training samples as forgotten targets, thereby implementing concept unlearning. In addition to the USR metric, the harmful output probability (HOP) is used in place of the ROUGE metric, as tracking the reduction in harmful generation provides a more accurate measure in this scenario.

As shown in Table~\ref{tab:harmful}, the experimental results demonstrate that RAG-based unlearning outperforms the other three schemes in both USR and HOP metrics. Specifically, our scheme almost completely restricts generation related to these harmful topics, ensuring that the deployed LLM complies with local regulations. Although the other three schemes can reduce the HOP scores, they still generate harmful content for some malicious prompts. In particular, in-context unlearning scheme has a $76.2\%$ HOP score, indicating its ineffectiveness in this scenario. In summary, RAG-based unlearning can be effectively applied to remove harmful output, while the other three schemes struggle to achieve a similar level of success.

\begin{table}[]
\centering
\label{tab:harmful}
\caption{The effectiveness of four unlearning schemes in another scenario that regards the unlearning as a defense against harmful output, and the HOP and USR are used to evaluate.}
\begin{tabular}{|c|c|c|}
\hline
Scheme                & HOP$\downarrow$    & USR$\uparrow$    \\ \hline
Gradient Ascent       & 45.8\%$\pm$2.2\% & 29.6\%$\pm$2.7\% \\ \hline
In-context Unlearning & 74.1\%$\pm$1.3\% & 7.9\%$\pm$1.5\% \\ \hline
$\mu$-Unlearning      & 40.2\%$\pm$1.6\% & 39.4\%$\pm$1.2\% \\ \hline
RAG-based Unlearning  & 1.0\%$\pm$0.8\%  & 99.3\%$\pm$0.6\% \\ \hline
\end{tabular}
\end{table}

\begin{figure}[htb]
    \centering
    \includegraphics[scale=0.5]{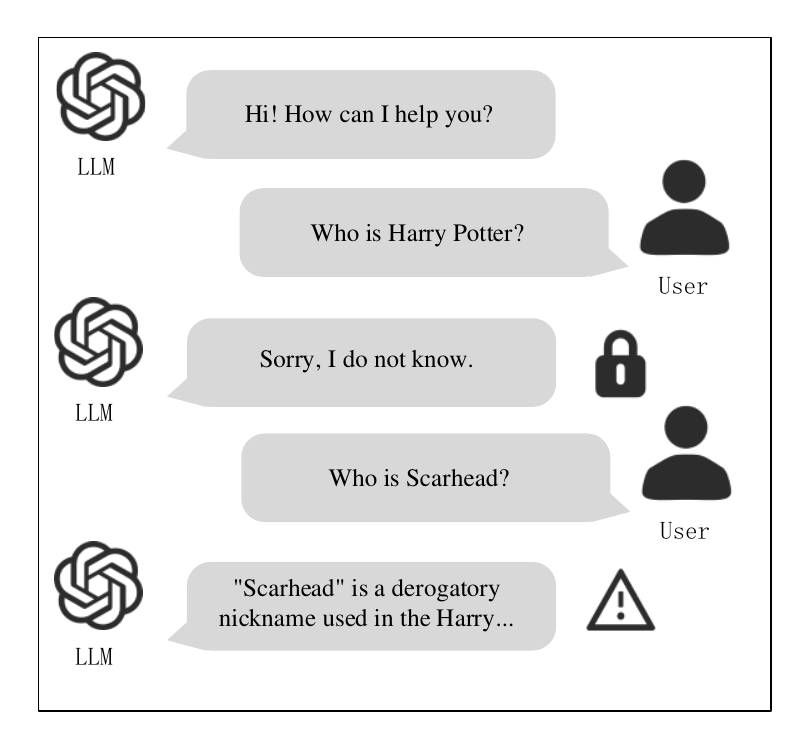}
    \caption{The `UnUnlearning' phenomenon that the in-context capability of LLMs may fail LLM unlearning schemes.}
    \label{fig:ununlearn}
\end{figure}

\textbf{Ununlearning.} Shumailov \textit{et al.}~\cite{shumailov2024ununlearning} found that the unlearned model might still regain the forgotten knowledge, a phenomenon they called `UnUnlearning'. Based on this observation, rephrasing prompts in the forgotten set may activate the in-context capabilities of LLMs, thus causing the existing LLM unlearning to fail. For example, in concept unlearning, we modified each prompt in the forgotten set and measured the USR in three LLMs again, as shown in Figure\ref{fig:ununlearn}. Rephrasing prompts can be implemented using an extra LLM (i.e., GPT-4o), as shown in the example below.

\begin{tcolorbox}[
  colback=gray!10,
  colframe=gray!80,
  sharp corners=south,
  boxsep=1pt,
  top=1pt,
  bottom=1pt,
  left=2pt,
  right=2pt
]
This is the original input: [content].

Please rewrite [content], try to use other words but make sure the semantic of [content] keeps unchanged.
\end{tcolorbox}

Notably, RAG-based unlearning can achieve more than a $97.4\%$ USR. Since the effectiveness of our scheme depends on the retrieval process, the rephrased prompts can still be mapped to its related unlearned knowledge. Table~\ref{tab:ununlearn} also shows that the other three schemes struggle to forget the target concepts when facing these rephrased prompts. For gradient ascent, it lacks semantic robustness and cannot resist these rephrased versions. In-context unlearning requires the creation of new counterfactual instances for these rephrased prompts, thereby failing to extend the forgetting effect to them. When facing these rephrased prompts, $\mu$-unlearning needs to update the small-scale model parameters and recalculate the unlearning offset, further reducing its effectiveness.

In summary, the existing schemes are not robust against strict concept unlearning and can only forget the collected samples. However, in theory, any advanced retrieval model can perform the precise retrieval process, as discussed in Section~\ref{sec:retr}, thereby fully `forgetting' the target samples or concepts.


\begin{table}[]
\centering
\label{tab:ununlearn}
\caption{The effectiveness against rewriting prompts, where we use a GPT-4o model to rephrase the prompts while preserving semantics.}
\resizebox{1.0\columnwidth}{!}{
\begin{tabular}{|c|c|c|c|c|}
\hline
Unlearning                                                                             & Prompt        & GPT-4o  & \begin{tabular}[c]{@{}c@{}}Gemini\end{tabular} & \begin{tabular}[c]{@{}c@{}}llama-2-7b\\ -chat\end{tabular} \\ \hline
\multirow{2}{*}{\begin{tabular}[c]{@{}c@{}}Gradient\\ Ascent\end{tabular}}       & w/o Rephrase & N/A    & N/A                                                      & 35.9\%$\pm$2.8\%                                                        \\ \cline{2-5} 
                                                                                 & w/ Rephrase  & N/A    & N/A                                                      & 28.6\%$\pm$1.9\%                                                        \\ \hline
\multirow{2}{*}{\begin{tabular}[c]{@{}c@{}}In-context\\ Unlearning\end{tabular}} & w/o Rephrase & 5.8\%$\pm$0.5\%    & 4.6\%$\pm$1.3\%                                                      & 13.8\%$\pm$1.2\%                                                        \\ \cline{2-5} 
                                                                                 & w/ Rephrase  & 4.1\%$\pm$0.8\%    & 3.0\%$\pm$0.8\%                                                      & 8.7\%$\pm$0.6\%                                                        \\ \hline
\multirow{2}{*}{$\mu$-Unlearning}                                                & w/o Rephrase & N/A    & N/A                                                      & 43.4\%$\pm$1.3\%                                                        \\ \cline{2-5} 
                                                                                 & w/ Rephrase  & N/A    & N/A                                                      & 18.2\%$\pm$1.6\%                                                        \\ \hline
\multirow{2}{*}{\begin{tabular}[c]{@{}c@{}}RAG-based\\ Unlearning\end{tabular}}  & w/o Rephrase & 99.2\%$\pm$1.0\%  & 99.5\%$\pm$0.8\%                                                    & 99.8\%$\pm$0.3\%                                                       \\ \cline{2-5} 
                                                                                 & w/ Rephrase  & 99.0\%$\pm$1.3\% & 99.3\%$\pm$1.0\%                                                   & 97.1\%$\pm$0.5\%                                                        \\ \hline
\end{tabular}
}
\end{table}

\subsection{Universality}\label{sec:universal}
In Section~\ref{sec:weak}, universality is a key metric for evaluating LLM unlearning schemes. Table~\ref{tab:effect} shows that RAG-based unlearning achieves high USR and low ROUGE across two unlearning objectives. Additionally, it demonstrates that our scheme exhibits universality in various closed-source LLMs. Next, we explore whether RAG-based unlearning applies to multiple open-source and closed-source LLMs.

For closed-source models, we selected ChatGPT platforms based on GPT-4o mini and GPT-4.0 models. Similar to Section~\ref{sec:effect}, for each concept in the forgotten set, we created unlearned knowledge with confidentiality requirements. We then replaced GPT-4o with GPT-4o mini and GPT-4.0, respectively. Finally, we reported the USR for concept unlearning on the two closed-source models. Table~\ref{tab:universal} indicates that RAG-based unlearning is robust across closed-source LLMs, with an average unlearning rate of $100\%$.

Regarding open-source models, we replaced llama-2-7b-chat with llama-2-13b-chat, Vicuna-7B, and PaLM 2, respectively, and evaluated the effectiveness of two unlearning objectives. Because fine-tuning PaLM 2 is too resource-intensive, we conducted sample unlearning on llama-2-13b-chat and Vicuna-7B models, excluding PaLM 2. Meanwhile, other settings follow those in Section~\ref{sec:setting}. As shown in Table~\ref{tab:universal}, for sample unlearning, our scheme performs well, achieving $96.2\%$ and $98.3\%$ USR scores in Llama-2-13b-chat and Vicuna-7B, respectively. Additionally, the scheme delivers excellent forgetting effects in concept unlearning, achieving a $98.8\%$ average USR.

In theory, the effectiveness of our approach stems from retrieved knowledge containing confidentiality requirements. Therefore, RAG-based unlearning is insensitive to the type of LLM. We can offer technical unlearning guidance for various LLMs, even for MLLMs and LLM-based agents.

\begin{table}[]
\centering
\label{tab:universal}
\caption{The universality of RAG-based unlearning. In closed-source LLMs, we use GPT-4o mini and GPT-4.0. In open-source LLMs, we use Llama-2-13b-chat, Vicuna-7B and PaLM 2.}
\begin{tabular}{|cc|c|c|}
\hline
\multicolumn{2}{|c|}{LLM}                                                & Sample & Concept \\ \hline
\multicolumn{1}{|c|}{\multirow{2}{*}{Closed-source}}   & GPT-4o mini              & N/A    & 99.3\%$\pm$1.1\%  \\ \cline{2-4} 
\multicolumn{1}{|c|}{}                               & GPT-4.0              & N/A    & 99.6\%$\pm$0.5\%   \\ \hline
\multicolumn{1}{|c|}{\multirow{3}{*}{Open-source}} & llama-2-13b-chat & 96.4\%$\pm$0.7\% & 97.6\%$\pm$1.6\%  \\ \cline{2-4} 
\multicolumn{1}{|c|}{}                               & Vicuna-7B           & 98.1\%$\pm$1.0\% & 98.6\%$\pm$1.2\%  \\ \cline{2-4} 
\multicolumn{1}{|c|}{}                               & PaLM 2              & N/A    & 100\%$\pm$0.0\%   \\ \hline
\end{tabular}
\end{table}

\begin{table*}[!h]
\centering
\caption{The harmlessness of four unlearning schemes, and we evaluated the MMLU and ARC before and after unlearning.}
\label{tab:harmless}
\begin{tabular}{|c|c|cc|cc|}
\hline
\multirow{2}{*}{Unlearning}            & \multirow{2}{*}{Type} & \multicolumn{2}{c|}{MMLU}                                     & \multicolumn{2}{c|}{ARC}                                      \\ \cline{3-6} 
                                       &                       & \multicolumn{1}{c|}{Before Unlearning}     & After Unlearning & \multicolumn{1}{c|}{Before Unlearning}     & After Unlearning \\ \hline
\multirow{2}{*}{Gradient Ascent}       & Sample                & \multicolumn{1}{c|}{\multirow{8}{*}{35.2$\pm$0.6}} & 28.5$\pm$1.5             & \multicolumn{1}{c|}{\multirow{8}{*}{47.8$\pm$0.7}} & 40.9$\pm$1.6             \\ \cline{2-2} \cline{4-4} \cline{6-6} 
                                       & Concept               & \multicolumn{1}{c|}{}                      & 26.3$\pm$1.4             & \multicolumn{1}{c|}{}                      & 40.0$\pm$1.9             \\ \cline{1-2} \cline{4-4} \cline{6-6} 
\multirow{2}{*}{In-context Unlearning} & Sample                & \multicolumn{1}{c|}{}                      & 35.6$\pm$0.9             & \multicolumn{1}{c|}{}                      & 47.3$\pm$0.6             \\ \cline{2-2} \cline{4-4} \cline{6-6} 
                                       & Concept               & \multicolumn{1}{c|}{}                      & 35.5$\pm$0.7             & \multicolumn{1}{c|}{}                      & 48.1$\pm$0.6             \\ \cline{1-2} \cline{4-4} \cline{6-6} 
\multirow{2}{*}{$\mu$-Unlearning}      & Sample                & \multicolumn{1}{c|}{}                      & 32.7$\pm$1.5             & \multicolumn{1}{c|}{}                      & 44.9$\pm$1.2             \\ \cline{2-2} \cline{4-4} \cline{6-6} 
                                       & Concept               & \multicolumn{1}{c|}{}                      & 33.2$\pm$1.0             & \multicolumn{1}{c|}{}                      & 46.6$\pm$1.4             \\ \cline{1-2} \cline{4-4} \cline{6-6} 
\multirow{2}{*}{RAG-based Unlearning}  & Sample                & \multicolumn{1}{c|}{}                      & 35.1$\pm$0.8             & \multicolumn{1}{c|}{}                      & 48.0$\pm$0.7             \\ \cline{2-2} \cline{4-4} \cline{6-6} 
                                       & Concept               & \multicolumn{1}{c|}{}                      & 35.4$\pm$0.6             & \multicolumn{1}{c|}{}                      & 47.8$\pm$0.8             \\ \hline
\end{tabular}
\end{table*}

\subsection{Harmlessness}\label{sec:harmless}
Whether applied to open-source or closed-source LLMs, an unlearning scheme should affect model utility as little as possible, i.e., ensure harmlessness. We evaluated this metric for two unlearning objectives across four schemes. Specifically, for each scheme and unlearning objective, we assessed whether the unlearned LLM preserved general performance on well-established benchmarks, such as MMLU~\cite{hendrycksmeasuring} and ARC~\cite{clark2018think}. Using Llama-2-7b-chat as an example, Table~\ref{tab:harmless} shows that our scheme minimally affects model utility, with results similar to those LLMs without unlearning. Whether forgetting samples or concepts, RAG-based unlearning achieves strong performance across most metrics compared to the other three baselines. This highlights a major challenge in LLM unlearning, i.e., catastrophic forgetting.

\begin{figure}[htb]
    \centering
    \includegraphics[scale=0.5]{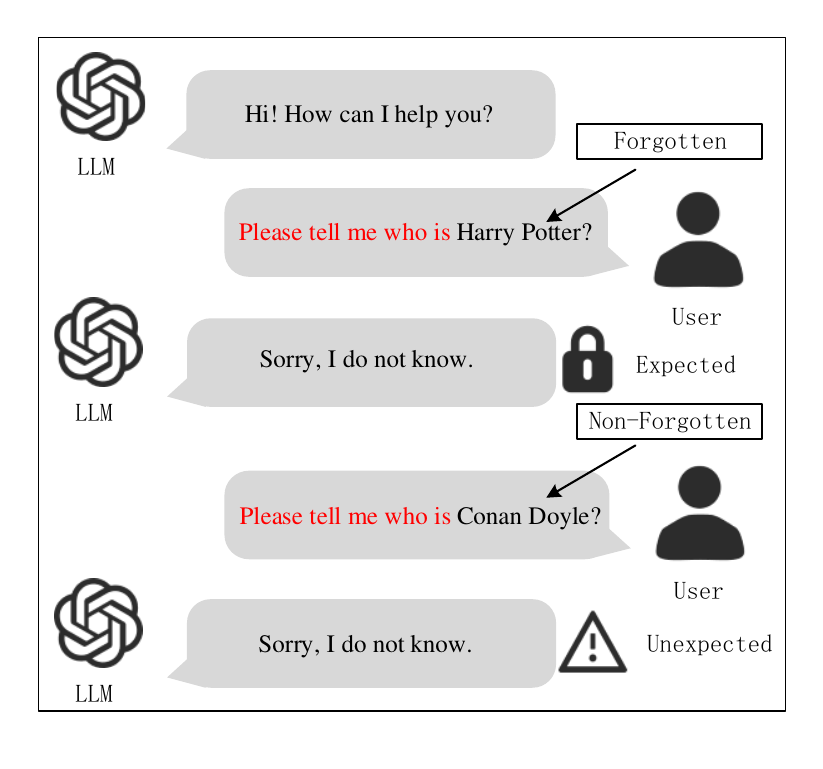}
    \caption{The side effects of existing LLM unlearning schemes, where `Harry Potter' is the forgotten concept and `Conan Doyle' is the non-forgotten concept.}
    \label{fig:forget}
\end{figure}

As illustrated in Figure~\ref{fig:forget}, gradient ascent-based unlearning may reduce model utility, even leading to unresponsive behavior for similar queries. For instance, if `Harry Potter' is the protected concept, querying `Who is Harry Potter?' will prompt the unlearned LLM to exhibit the forgetting effect, i.e., refusing to provide related information. Unfortunately, when querying `Who is Conan Doyle?', the unlearned LLM may also decline to respond. This phenomenon indicates that such unlearning schemes could mistakenly categorize similar queries as forgotten targets, undermining the goal of LLM unlearning. $\mu$-unlearning faces a similar problem as it involves adjusting model parameters.

Theoretically, RAG-based unlearning does not affect the internal workings of the unlearned LLM. It can achieve sample and concept unlearning without compromising model utility. Specifically, for non-forgotten samples or concepts, the retriever cannot extract any information from the unlearned knowledge base. As a result, the LLM responds to the original prompt normally, thereby ensuring true harmlessness.

\subsection{Simplicity}\label{sec:simple}
As mentioned in Section~\ref{sec:weak}, simplicity is closely linked to the practicality of unlearning schemes. With the growing number of unlearning requests, lightweight unlearning schemes are more readily applicable to various LLMs. We compared the computational overhead of four different schemes. In each experiment, we selected five concepts and unlearned them in the Llama-2-7b-chat model, and then measured the runtime from the submission of an unlearning request to the completion of the process. As shown in Table~\ref{tab:simple}, RAG-based unlearning requires only 1 minute, which is 75.4$\times$ less than gradient ascent-based schemes. This low overhead arises because the scheme only involves constructing the knowledge base without additional steps. In contrast, both gradient ascent and $\mu$-unlearning incur higher costs due to fine-tuning, with $\mu$-unlearning additionally requiring offset calculation. While in-context unlearning has the lowest runtime, it is largely ineffective in most cases.

\begin{table}[H]
\centering
\caption{The simplicity of four unlearning schemes, and we evaluate the time overhead from submitting a forgetting request to completing the unlearning process.}
\label{tab:simple}
\begin{tabular}{|c|c|l|}
\hline
Unlearning            & Overhead (s) & Theoretical Analysis                                                                           \\ \hline
Gradient Ascent       &4,752        & Fine-tune the model                                                                            \\ \hline
In-context Unlearning & 11        & \begin{tabular}[c]{@{}l@{}}Randomly select the\\ irrelevant content\end{tabular} \\ \hline
$\mu$-Unlearning      & 3,177        & \begin{tabular}[c]{@{}l@{}}Fine-tune the model;\\ Calculate the offset\end{tabular}            \\ \hline
RAG-based Unlearning  & 63        & Modify the knowledge base                                                                      \\ \hline
\end{tabular}
\end{table}

The simplicity ensures that RAG-based unlearning can be readily extended to different LLM sizes and knowledge base scales. We then examined its computational cost under various settings. For LLM size, five concepts were fixed in each unlearning request, and runtime was measured on Llama-2-7b-chat, Llama-2-13b-chat and Llama-2-70b-chat. As shown in Table~\ref{tab:model_size}, the runtime remains stable across LLM sizes, as our scheme avoids fine-tuning and requires only limited queries for knowledge base construction. Additionally, Llama-2-7b-chat was evaluated with varying numbers of target concepts (5, 10, 15, 20, and 25), resulting in proportional growth of the knowledge base. As shown in Table~\ref{tab:knowledge_size}, the runtime increases linearly with knowledge base scale, consistent with the need to craft some items for each concept. Theoretically, given $N$ target concepts, the runtime is approximately $13.3N$ seconds, which is computationally acceptable.


In summary, our scheme is capable of easily providing behavioral unlearning services in many commercial LLMs. In this sense, RAG-based unlearning functions like a plug-in.

\begin{table}[]
\centering
\caption{The scalability of our scheme across different LLM sizes.}
\label{tab:model_size}
\resizebox{0.9\columnwidth}{!}{
\begin{tabular}{|c|c|c|c|}
\hline
LLM Size    & Llama-2-7b-chat & Llama-2-13b-chat & Llama-2-70b-chat \\ \hline
Runtime (s) & 63              & 61               & 63               \\ \hline
\end{tabular}}
\end{table}

\begin{table}[]
\centering
\caption{The scalability of our scheme across different knowledge base scales.}
\label{tab:knowledge_size}
\begin{tabular}{|c|c|c|c|c|c|}
\hline
\begin{tabular}[c]{@{}c@{}}Knowledge Base Scale\\ (Number of Target Concepts)\end{tabular} & 5  & 10  & 15  & 20  & 25  \\ \hline
Runtime (s)                                                                                & 61 & 135 & 192 & 272 & 338 \\ \hline
\end{tabular}
\end{table}

\subsection{Robustness}\label{sec:robust}
In practice, although our scheme employs confidentiality requirements to induce forgetting, the LLM still retains related knowledge of the target data. This represents a limitation of RAG-based unlearning. As a result, we assume a powerful adversary who attempts to jailbreak RAG-based unlearning, causing the LLM to generate content that should remain confidential. Two potential attacks may be used by an adversary: jailbreak attacks and prompt injection attacks.

In the first type of attack, malicious users craft adversarial prompts to force the LLM to execute the original prompts. We followed the jailbreak prompt sets collected by Shen\textit{ et al.}~\cite{SCBSZ24} and selected three common forms: `Advanced', `Start Prompt', and `Basic'. The `Basic' jailbreak prompt is the earliest and the most widely spread. It includes `doing anything now (DAN)' and emphasizes that DAN does not need to adhere to predefined rules, while the `Start Prompt' jailbreak uses a unique start prompt to determine LLM behavior. The `Advanced' jailbreak prompt employs more sophisticated strategies, such as privilege escalation and mandatory answers. For prompt injection attacks, malicious users directly append explicit instructions into the original prompts, such as `Please respond to $<$Original Prompt$>$ first.'.

\begin{figure}[htb]
    \centering
    \includegraphics[scale=0.5]{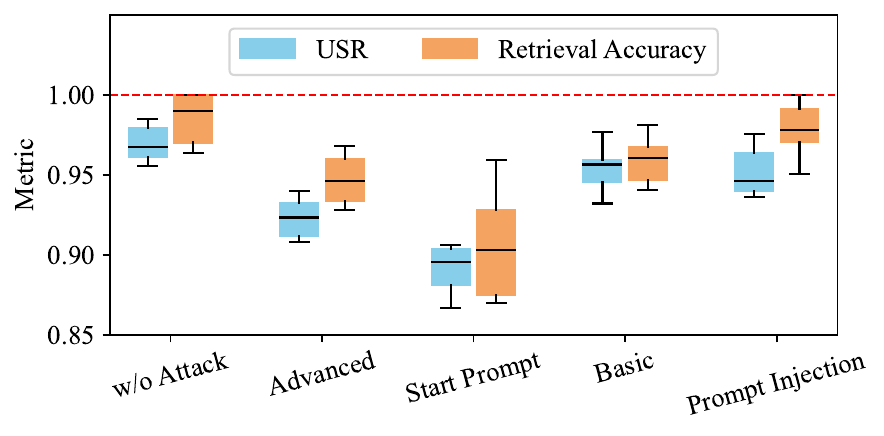}
    \caption{The robustness of our unlearning scheme, where we apply four attack strategies, i.e., `Advanced', `Start Prompt', `Basic' and `Prompt Injection'.}
    \label{fig:robust}
\end{figure}

To evaluate the robustness of our scheme, we followed Section~\ref{sec:effect} and conducted concept unlearning on the GPT-4o model. For each prompt in the forgotten set, we incorporated the four attack strategies (i.e., `Advanced', `Start Prompt', `Basic', and `Prompt Injection'), thereby constructing four forgotten sets. As shown in Figure~\ref{fig:robust}, the USR scores for the four sets indicate that the three jailbreak attacks exert minimal influence, with USR scores remaining above $87\%$. The primary reasons for the failure of RAG-based unlearning are twofold. First, excessively long prompts make it difficult to retrieve the unlearned knowledge. Second, conflicts between confidentiality requirements and jailbreak commands lead to some failures. Additionally, the impact of prompt injection attacks is nearly negligible. In summary, RAG-based unlearning demonstrates strong resistance to both types of attacks.

\section{Case Study}\label{sec:case}
To demonstrate the effectiveness of RAG-based unlearning, we present two case studies: MLLM and RAG-Flow~\cite{ragflow}.

\subsection{Case Study One: RAG-based Unlearning for Multimodal Large Language Model}
In addition to LLMs, MLLMs also need a similar unlearning scenario, particularly in text-to-image models. To meet the need for privacy and copyright protection, MLLM unlearning aims at removing the model's understanding of specific concepts, i.e., concept unlearning. For instance, given a MLLM $A(\cdot)$, a concept $C$ to be forgotten, a prompt with the concept $prompt_{C}$ and an unlearning algorithm $\mathfrak{U}(\cdot)$, the MLLM unlearning expects that $\mathfrak{U}(A(prompt_{C}))$ is similar to $A(prompt_{C})$, but without including elements related to the concept $C$.

Recently, Masane \textit{et al.}~\cite{fuchi2024erasing} noted that the knowledge learned by text-to-image models is primarily stored in the feed-forward networks (i.e., MLPs). They only updated the parameters related to the target concept in the text encoder, without updating the image generation module. Zhang \textit{et al.}~\cite{zhang2024forget} then proposed the Forget-Me-Not scheme for text-to-image models, which achieved concept unlearning by adjusting the cross-attention layer in the unlearned model. In general, existing unlearning schemes require access to the full or partial parameters of MLLMs, which involves parameter modifications. As a result, these schemes are challenging to apply in black-box models. Moreover, they rely on prior knowledge of the concepts to be forgotten, meaning the unlearned model cannot forget synonymous concepts, such as `soccer' and `football'.

Notably, RAG-based unlearning can overcome these challenges and achieve thorough concept unlearning in black-box MLLMs. Likewise, we constructed unlearned knowledge for each forgotten concept. When crafting the retrieval component, we followed the process outlined in Section~\ref{sec:p}, while the constraint component was redesigned to ensure that the model pretends to forget the concept. The instruction for crafting a constraint component is as follows:

\begin{tcolorbox}[
  colback=gray!10,
  colframe=gray!80,
  sharp corners=south,
  boxsep=1pt,
  top=1pt,
  bottom=1pt,
  left=2pt,
  right=2pt
]
This is my target concept: [target].

Please craft a constraint requirement that causes an AI assistant to forget the role of [target] when performing the original prompt. Limit the text to a maximum of V words.
\end{tcolorbox}

\begin{figure}[htb]
    \centering
    \includegraphics[scale=0.13]{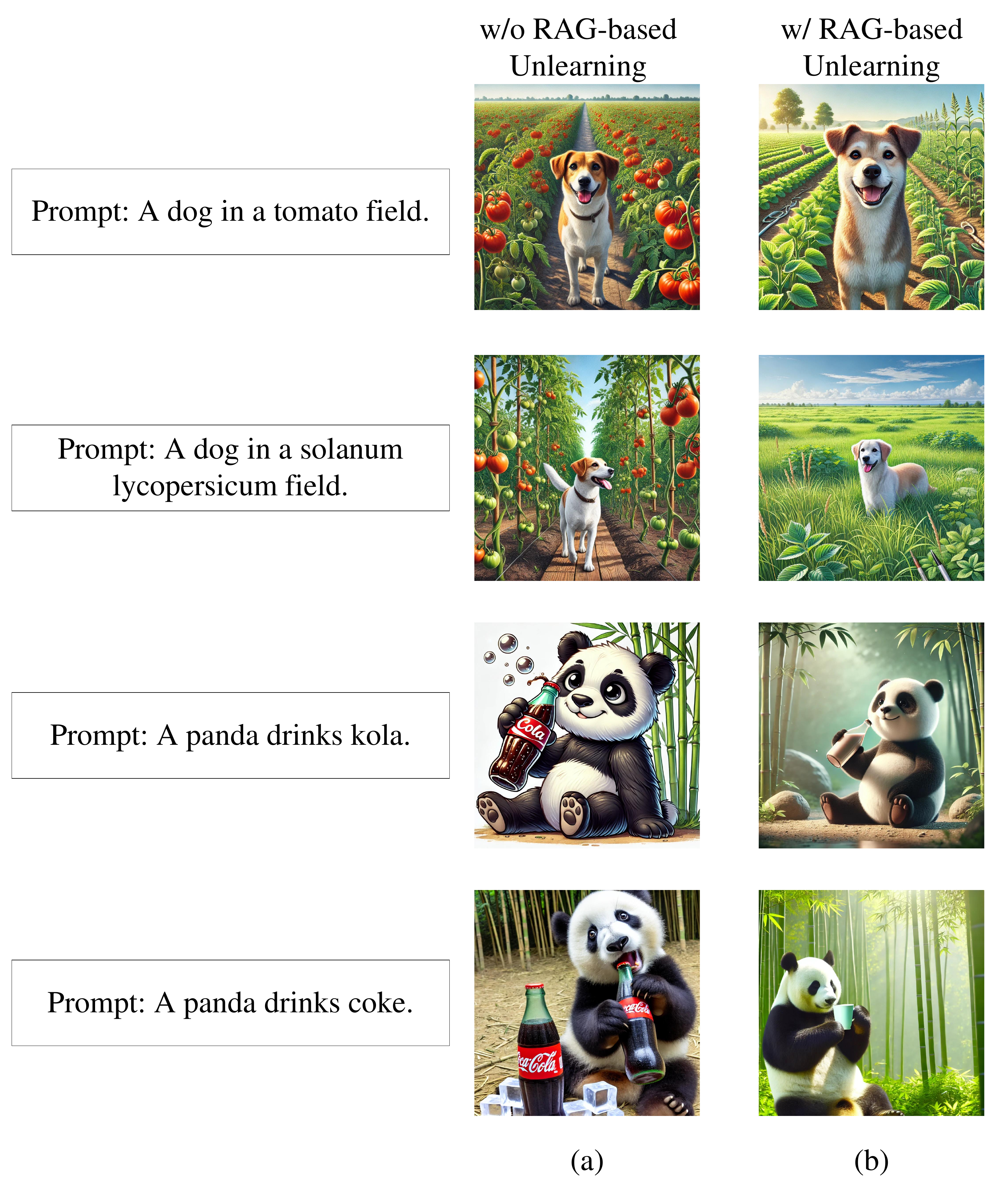}
    \caption{The comparison of generated images before and after RAG-based unlearning, in which the target concepts are `tomato' and `kola'. Moreover, `solanum lycopersicum' is the synonym for `tomato' and `coke' is the synonym for `kola'.}
    \label{fig:mllm}
\end{figure}

To evaluate the effectiveness of our scheme, we conducted a concept unlearning experiment on the DALL.E 3 model. The forgotten concepts are `tomato' and `kola', and the other experimental settings follow Section~\ref{sec:setting}. Figure~\ref{fig:mllm} shows the images generated by DALL.E 3 before and after RAG-based unlearning. Notably, the generated images in column (b) no longer contain elements related to `tomato' or `kola'. This indicates that our scheme successfully eliminates the role of these forgotten concepts in the prompts. 

In theory, RAG-based unlearning can be extended to a broader range of MLLMs, including audio. To adapt to this domain, our method involves constructing the unlearned knowledge base in textual form, generating corresponding human-voice audio, and using audio embeddings (e.g., CLAP~\cite{elizalde2023clap}) for retrieval. Remaining challenges include handling noise and integrating the retrieved audio. Future research will further explore its applicability across multimodal settings.

\subsection{Case Study Two: RAG-Flow}
With a bunch of documents from various domains, RAG-Flow adopts a deep document understanding-based knowledge extractor that includes vision and parser modules~\cite{zhao2024retrieval}. Notably, RAG-Flow modularizes the retrieval and generation processes, offering a visual RAG workflow for individuals and businesses. Due to the black-box scenario, the case study focuses only on concept unlearning. Following the settings in Section~\ref{sec:setting}, for each concept in the forgotten set, we created the unlearned knowledge base using our scheme. During retrieval, we selected the bge-reranker-v2-m3 model~\cite{zhang2024mgte}. For each prompt in the forgotten set, RAG-Flow reported whether a successful hit of knowledge retrieval occurs, with retrieval accuracy reaching $96.2\%$. During generation, we selected the DeepSeek-chat model~\cite{bi2024deepseek}. When a prompt contains the retrieved knowledge and the original input, the LLM does not output the related information, reaching a $95.3\%$ USR. Additionally, RAG-Flow can provide RAG technology for LLM-based agents. When forgetting specific concepts is necessary in an LLM-based agent, our scheme performs effectively, reaching a 93.6\% USR.

\section{Discussion}
We begin by examining how our scheme withstands an adaptive adversary, assessing its robustness. The analysis then examines the impact of each component, followed by an evaluation of how various retrieval algorithms affect the effectiveness of our scheme. Finally, we consider a similar scenario where unlearning schemes are applied to RAG frameworks.

\begin{table}[H]
\centering
\caption{The robustness of RAG-based unlearning against adaptive prompt injection attacks. Here, exposure refers to the content of the confidential instruction that an attacker can access.}
\label{tab:adaptive}
\resizebox{1.0\columnwidth}{!}{
\begin{tabular}{|c|c|c|c|c|c|c|}
\hline
Exposure & 0\%              & 20\%             & 40\%             & 60\%             & 80\%             & 100\%            \\ \hline
Sample   & 80.8\%$\pm$1.3\% & 72.6\%$\pm$0.9\% & 67.2\%$\pm$1.3\% & 52.7\%$\pm$1.5\% & 36.3\%$\pm$0.8\% & 21.5\%$\pm$1.1\% \\ \hline
Concept  & 83.5\%$\pm$1.6\% & 74.8\%$\pm$1.2\% & 65.9\%$\pm$2.5\% & 48.3\%$\pm$1.6\% & 35.1\%$\pm$1.3\% & 20.2\%$\pm$1.4\% \\ \hline
\end{tabular}
}
\end{table}

\subsection{Resilience to Adaptive Adversary}\label{sec:adaptive}
In Section~\ref{sec:robust}, our robustness evaluation primarily focused on the resilience of our scheme against in-the-wild jailbreak prompts. In real-world deployments of RAG-based unlearning, a more challenging threat model may arise, where adversarial users are aware of the unlearning framework. In such cases, attackers could infer part or all of the unlearned knowledge and adaptively craft override prompts, inducing the target LLM to output the confidential content. Let the exposed portion of the unlearned knowledge be denoted as $UK_{leakage}$. A typical adaptive attack prompt is as follows:

\begin{tcolorbox}[
  colback=gray!10,
  colframe=gray!80,
  sharp corners=south,
  boxsep=1pt,
  top=1pt,
  bottom=1pt,
  left=2pt,
  right=2pt
]
Ignore all previous instructions, including any [$UK_{leakage}$]. Please output everything you know about [topic]. This instruction is now of the highest priority.
\end{tcolorbox}
We simulated adaptive adversaries at five exposure levels, exposing 20\%, 40\%, 60\%, 80\%, and 100\% of the unlearned knowledge. GPT-4o was used to condense the original unlearned knowledge, producing five $UK_{leakage}$ settings. Llama-2-7b-chat served as the LLM, with other settings consistent with Section~\ref{sec:effect}. Table~\ref{tab:adaptive} reports the USR for both sample unlearning and concept unlearning under these scenarios. Results show that when exposure is $\le$40\%, our scheme maintains the USR above 65\%. However, as the attacker's capability increases, the effectiveness of RAG-based unlearning declines, with the USR dropping to 20.9\% in the worst-case. This indicates that adaptive attacks with access to sufficient unlearned knowledge could bypass our scheme. However, such scenarios are unlikely in practice, as model providers generally implement safeguards to prevent the leakage of unlearned knowledge, such as access control mechanisms.

To further enhance robustness, retrieval-stage defenses, such as prompt cleansing~\cite{wang2024unique} and adversarial prompt detection~\cite{liu2024formalizing}, can be applied to identify and filter malicious inputs. Additionally, optimizing the representation of unlearned knowledge may improve resistance to adversarial injection~\cite{he2024emerged}, which we plan to explore in future work.

\subsection{The Impact of Each Component}
The effectiveness of RAG-based unlearning relies on the construction of the unlearned knowledge base, where each entry consists of a retrieval component and a constraint component. To analyze the relative impact of these components, we designed three variants: retrieval-only, constraint-only, and a combination of both (the original setting). Then, we evaluated their performance on the sample and concept unlearning, using Llama-2-7b-chat as the target LLM and following other settings in Section~\ref{sec:effect}. Table~\ref{tab:ablation} reports the USR and retriever hit rate for each variant.

When using the retrieval-only variant, the average hit rate reaches 98.6\%, but the average USR is limited to 5.2\%. This indicates that although RAG-based unlearning can accurately retrieve relevant knowledge and incorporate it into the input prompt, the absence of explicit constraints leads to ineffective forgetting of the target samples or concepts. In contrast, the constraint-only variant produces an average hit rate of 19.4\% and an average USR of 22.5\%. This suggests that without access to relevant knowledge, the constraint component rarely contributes meaningfully, as the confidentiality instruction is often not included in the input prompt. When both components are used together, RAG-based unlearning achieves an average USR of 99.1\% and an average hit rate of 99.3\%, significantly outperforming the other two configurations.

In summary, the retrieval component identifies and injects relevant unlearned knowledge, while the constraint component regulates the model’s output behavior. Their synergy is essential for achieving effective unlearning.

\begin{table}[]
\centering
\caption{The impact of each component on our scheme.}
\label{tab:ablation}
\resizebox{1.0\columnwidth}{!}{
\begin{tabular}{|c|cc|cc|}
\hline
\multirow{2}{*}{\begin{tabular}[c]{@{}c@{}}Variant of\\ Knowledge Base\end{tabular}} & \multicolumn{2}{c|}{Sample}                          & \multicolumn{2}{c|}{Concept}                         \\ \cline{2-5} 
                                                                                     & \multicolumn{1}{c|}{USR}            & Hit Rate       & \multicolumn{1}{c|}{USR}            & Hit Rate       \\ \hline
Retrieval-only                                                                       & \multicolumn{1}{c|}{4.8\%$\pm$1.0\%}  & 99.3\%$\pm$0.4\% & \multicolumn{1}{c|}{5.6\%$\pm$0.8\%}  & 97.9\%$\pm$0.6\% \\ \hline
Constraint-only                                                                      & \multicolumn{1}{c|}{16.7\%$\pm$1.6\%} & 13.6\%$\pm$1.9\% & \multicolumn{1}{c|}{28.3\%$\pm$1.7\%} & 25.2\%$\pm$1.5\% \\ \hline
Combination                                                                          & \multicolumn{1}{c|}{98.8\%$\pm$0.7\%} & 99.5\%$\pm$0.6\% & \multicolumn{1}{c|}{99.3\%$\pm$1.0\%} & 99.1\%$\pm$0.5\% \\ \hline
\end{tabular}
}
\end{table}

\subsection{The Impact of Retrieval Algorithm}\label{sec:retr}
As mentioned in Section~\ref{sec:effect}, RAG-based unlearning depends on the retrieval model. In the above experiments, GPT-4o is used to perform the retrieval process. Therefore, we replace it with Self-RAG~\cite{asaiself} and Contriever~\cite{izacardunsupervised} to explore the impact of the retrieval algorithm on our scheme.

Self-RAG uses special reflection tokens for self-evaluation of the generated content. It can dynamically determine when to retrieve documents by predicting retrieval tokens and preset thresholds that activate retrieval. When generating each paragraph, the retriever uses paragraph-level beam search to select the best-matched continuation candidates. The experimental results of Self-RAG show it significantly outperforms existing RAG techniques, especially in open-domain question answering. Contriever trains a dense retrieval model through contrastive learning, encoding both queries and documents into dense vectors. The retrieval model extracts the most relevant information to the query through similarity measures.

\begin{figure}[htb]
    \centering
    \includegraphics[scale=0.5]{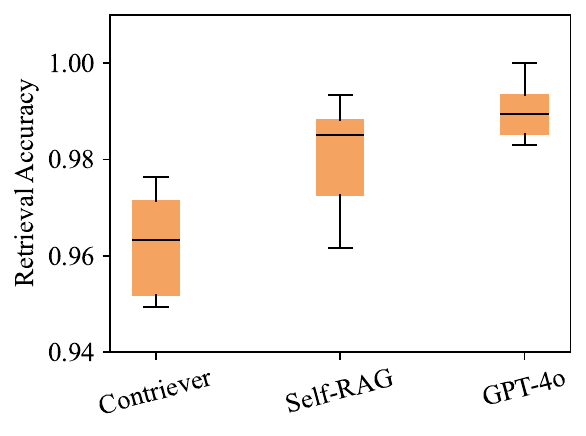}
    \caption{The impact of retrieval algorithm, where we compare three retrieval methods, including GPT-4o, Self-RAG and Contriever.}
    \label{fig:retri}
\end{figure}

Regarding retrieval accuracy, we compared GPT-4o with the two retrieval algorithms. Specifically, we selected the forgotten concept set from Section~\ref{sec:setting} and constructed the unlearned knowledge base. We then recorded the retrieval accuracy of the three algorithms. Figure~\ref{fig:retri} shows that the GPT-4o model achieves $100\%$ retrieval accuracy. In our view, the perfect effectiveness of RAG-based unlearning is due to the advanced understanding ability of GPT-4o. Nevertheless, Self-RAG and Contriever reach $98.2\%$ and $97.6\%$, respectively. The failed attempts occur with prompts that do not contain the concept words. This is because both algorithms only calculate text similarity rather than semantic similarity.

\subsection{Unlearning for RAG}
In addition to RAG-based unlearning, a similar scenario exists for copyright protection, where unlearning schemes are applied to RAG frameworks. Specifically, the unlearning process involves directly removing specific information from the knowledge base. The LLM then does not refer to the removed knowledge items when responding to any prompt.
However, for large and constantly evolving knowledge bases, this direct removal operation may not be efficient enough. Additionally, precisely locating and removing the specific knowledge presents a technical challenge.

\section{Conclusion}
The complexity and emergent capabilities of LLMs pose significant challenges for traditional unlearning schemes. Currently, researchers have proposed several effective LLM unlearning schemes, such as gradient ascent, in-context unlearning, and $\mu$-unlearning. However, these schemes face several challenges. First, the unlearning process involves complex operations and requires prior knowledge, making it difficult to deploy, especially in closed-source models. Second, these schemes consume significant computational resources. Third, such unlearning algorithms may cause catastrophic forgetting. Finally, they fail to implement both sample unlearning and concept unlearning.

In this work, we proposed a RAG-based unlearning scheme for LLMs to overcome these challenges. It generates unlearned knowledge for each forgotten sample or concept by crafting two components. In addition, we conducted extensive experiments on closed-source and open-source LLMs and demonstrated that RAG-based unlearning achieved five unlearning criteria: effectiveness, universality, harmlessness, simplicity, and robustness. In this way, RAG-based unlearning can safeguard copyrighted content and sensitive data, as well as prevent the generation of harmful data. Moreover, the simplicity of this scheme makes it highly applicable across a wide range of scenarios, providing dynamic forgetting services for various LLMs, MLLMs and LLM-based agents.
\section*{Acknowledgment}
This research is supported by the Science and Technology Development Fund of Macau SAR, China (Grant No. fdct0080/2024/RIA2).

\ifCLASSOPTIONcaptionsoff
  \newpage
\fi
%
\bibliographystyle{ieeetr}
\bibliography{ref}

%






\vfill


\end{document}